\documentclass{emulateapj}
\usepackage{graphicx} \usepackage{color} \usepackage{natbib}

\usepackage{amsmath,amssymb,amsfonts}
\usepackage{epsf}
\usepackage{epsfig}
\usepackage{longtable}

\newcommand{\aanda}{A\&A}
\newcommand{\nucpha}{Nucl. Phys. A}
\newcommand{\phrep}{Phys. Rep.}
\newcommand{\cqg}{Classical and Quantum Gravity}
\newcommand{\arxiv}{arXiv}
\newcommand{\ppnucph}{Progress in Particle and Nuclear Physics}
\newcommand{\lrr}{Living Rev. Rel.}
\newcommand{\sci}{Science}
\newcommand{\ptp}{Prog. Theor. Phys.}
\newcommand{\nar}{NewAR}

\shorttitle{Systematics of dynamical mass ejection, nucleosynthesis, and radioactively powered electromagnetic signals}
\shortauthors{Bauswein, Goriely, Janka}

\begin{document}

\title{Systematics of dynamical mass ejection, nucleosynthesis, and radioactively powered electromagnetic signals from neutron-star mergers}

\author{A.~Bauswein\altaffilmark{1}, S.~Goriely\altaffilmark{2}, H.-T.~Janka\altaffilmark{1}}

\altaffiltext{1}{Max Planck Institute for Astrophysics, Karl-Schwarzschild-Str.~1, 85748 Garching, Germany}
\altaffiltext{2}{Institut d'Astronomie et d'Astrophysique, Universit\'e Libre de Bruxelles, C.P. 226, B-1050 Brussels, Belgium}

\begin{abstract}

We investigate systematically the dynamical mass ejection, r-process
nucleosynthesis, and properties of electromagnetic counterparts of
neutron-star (NS) mergers in dependence on the uncertain properties of the
nuclear equation of state (EoS) by employing 40 representative,
microphysical high-density EoSs in relativistic, hydrodynamical
simulations. The crucial parameter determining the ejecta mass is the radius $R_{1.35}$ of a 1.35\,$M_{\odot}$ NS. NSs with smaller 
$R_{1.35}$ (``soft'' EoS) eject systematically higher masses. These range
from $\sim$$10^{-3}\,M_\odot$ to $\sim$$10^{-2}\,M_\odot$ for 1.35-1.35~$M_{\odot}$ binaries
and from $\sim$$5\times 10^{-3}\,M_\odot$ to
$\sim$$2\times 10^{-2}\,M_\odot$ for
1.2-1.5\,$M_\odot$ systems (with kinetic energies between $\sim$$5\times 10^{49}$\,erg and
$10^{51}$\,erg). Correspondingly, the bolometric peak luminosities of the optical
transients of symmetric (asymmetric) mergers vary between $3\times
10^{41}$\,erg\,s$^{-1}$ and $14\times 10^{41}$\,erg\,s$^{-1}$ ($9\times
10^{41}$\,erg\,s$^{-1}$ and $14.5\times 10^{41}$\,erg\,s$^{-1}$) on
timescales between $\sim$2\,h and $\sim$12\,h. If these
signals with absolute bolometric magnitudes from $-$15.0 to $-$16.7 are
measured, the tight correlation of their properties with those of the
merging NSs might provide valuable constraints on the high-density EoS. The r-process nucleosynthesis exhibits
a remarkable robustness independent of the EoS, producing a nearly solar abundance pattern above mass number 130. By the r-process content of the Galaxy and the average production per event the Galactic merger rate is limited to $4\times
10^{-5}$\,yr$^{-1}$ ($4\times 10^{-4}$\,yr$^{-1}$) for a soft (stiff) NS
EoS, if NS mergers are the main source of heavy r-nuclei. The production
ratio of radioactive $^{232}$Th to $^{238}$U attains a stable value of
1.64--1.67, which does not exclude NS mergers as potential sources of
heavy r-material in the most metal-poor stars.

\end{abstract}


\section{Introduction}
Neutron star (NS) merger events are among the most promising 
candidates for the first direct measurement of a gravitational-wave
signal with the upcoming Advanced LIGO and VIRGO interferometric
instruments~\citep{2006CQGra..23S.635A,2010CQGra..27h4006H}, and they are considered as likely
origin of short gamma-ray bursts and their afterglows as a consequence
of ultrarelativistic, collimated outflows~\citep[see e.g.][]{2006ApJ...650..261S,2007PhR...442..166N,2011NewAR..55....1B,2011ApJ...734...96K,2012ApJ...756..189F}.
Moreover, they are possible sources of different kinds of electromagnetic signals in the precursor of the merging and in its aftermath as a consequence of magnetohydrodynamical effects, magnetospheric interactions, relativistic matter outflows, or NS crust phenomena~\citep{1996A&A...312..937L,1996ApJ...471L..95V,1998ApJ...507L..59L,2001MNRAS.322..695H,2010ApJ...723.1711T,2011ApJ...734L..36S,2011Natur.478...82N,2012PhRvL.108a1102T,2012arXiv1209.5747K,2013ApJ...763L..22Z,2013arXiv1301.0439G,2012ApJ...755...80P,2012ApJ...757L...3L,2013arXiv1301.7074P,2012ApJ...746...48M}.
Thermal emission produced by hot ejecta gas, for example, may
cause potentially observable optical transients~\citep{1998ApJ...507L..59L,2005astro.ph.10256K,2010MNRAS.406.2650M,2012ApJ...746...48M}, and the 
interaction of the ejecta cloud with the circumstellar medium is 
expected to create radio flares that might be detectable for periods
of years~\citep{2011Natur.478...82N,2012arXiv1204.6242P,2012arXiv1204.6240R}.
Observations of such signals could help to pinpoint the exact celestial
locations of NS mergers (thus, e.g., supporting the analysis of data
taken by gravitational-wave detectors), and repeated measurements of
signals that can be unambiguously linked to NS mergers would help to
constrain the still highly uncertain rate of such events in the local
universe.

During the merging of two NSs a small fraction of the system mass, 
typically 0.1--1 per cent, can become gravitationally unbound and 
can be ejected on the dynamical timescale of milliseconds~\citep{1997A&A...319..122R,1999A&A...341..499R,2000A&A...360..171R,2001A&A...380..544R,2007A&A...467..395O,2012arXiv1204.6240R,2012arXiv1204.6242P,2012arXiv1210.6549R,2012arXiv1212.0905H}. Because such material is
likely to possess a high neutron excess, it has been proposed as a 
possible site for the creation of the heaviest, neutron-rich
elements, which are formed by the rapid neutron capture process
(r-process)~\citep{1977ApJ...213..225L,1989Natur.340..126E}
(similarly, also NS-black hole mergers were suggested as sources of r-process matter~\citep{1974ApJ...192L.145L,1976ApJ...210..549L}). The radioactive
decay of these freshly synthesized r-process nuclei should heat the ejecta
and thus lead to an optical transient~\citep{1998ApJ...507L..59L,2005astro.ph.10256K,2010MNRAS.406.2650M,2011ApJ...736L..21R,2011ApJ...738L..32G}. 
The properties of such events depend on the fraction of the 
material that can be converted to radioactive species. Moreover,
the peak luminosity, the timescale to reach the emission peak, and the effective
temperature at the radiation maximum, as well as the radio brightness
that accompanies the deceleration of the expelled gas during its 
coasting in the
stellar environment, depend sensitively on the ejecta mass and expansion
velocity. Detailed hydrodynamical merger models are needed to calculate
these quantities and to determine the nucleosynthesis conditions
in the unbound material.

Concerning their role as sources of heavy elements
binary NSs collisions have recently moved into 
the focus of interest because the astrophysical sources of the r-process
elements have not been identified yet and core-collapse supernova 
simulations continue to be unable to yield the extreme conditions
for forming the heaviest neutron-rich nuclei~\citep{2007A&A...467.1227A,2008ApJ...676L.127H,2008A&A...485..199J,2010ApJ...722..954R,2010PhRvL.104y1101H,2010A&A...517A..80F,2011ApJ...726L..15W,2011A&A...526A.160A,2011PhRvC..83d5809A}. 
(For reviews on r-process
nucleosynthesis and an overview of potential sites, ~\citep[see e.g.][]{2007PhR...450...97A,2011PrPNP..66..346T,2011PhRvL.106t1104B,2012ApJ...750L..22W}.) 
In contrast to the situation for supernovae, investigations with growing
sophistication have confirmed NS merger ejecta as viable sites 
for strong r-processing~\citep{1999ApJ...525L.121F,2005NuPhA.758..587G,2007PhR...450...97A,2010MNRAS.406.2650M,2011ApJ...736L..21R,2011ApJ...738L..32G,2012arXiv1206.2379K}. 

However, despite this promising situation a variety of aspects need to
be clarified before the question can be answered whether NS mergers are
a major source or even the dominant source of heavy r-process elements. 
On the one hand the merger rate and its evolution during the Galactic
history are still subject to considerable uncertainties (see e.g.,~\citet{2010CQGra..27q3001A} for a compilation of recent estimates), and it is unclear whether NS mergers can
explain the early enrichment of the Galaxy by r-process elements as
observed in metal-deficient stars~\citep{2004A&A...416..997A}.
On the other hand it remains to be determined how much mass is 
ejected in merger events depending on the binary parameters and, 
in particular, depending on the incompletely known properties of the 
equation of state (EoS) of NS matter. It also needs to be
understood which fraction of the ejecta is robustly converted to 
r-process material and whether the final abundances are always 
compatible with the solar element distribution, which agrees amazingly
well with the r-process abundance pattern in metal-poor stars for atomic
numbers $Z~\sim 55$--90~\citep[see e.g.][]{2008ARA&A..46..241S}.

Newtonian as well as relativistic studies showed that the 
mass ratio has a significant effect on the amount of matter that can
become unbound~\citep{1999ApJ...527L..39J,1999A&A...341..499R,2000A&A...360..171R,2001A&A...380..544R,2007A&A...467..395O,2011ApJ...736L..21R,2011ApJ...738L..32G,2012arXiv1204.6242P,2012arXiv1204.6240R,2012arXiv1206.2379K,2012arXiv1210.6549R,2012arXiv1212.0905H}. 
Such investigations, however,
were performed only with a few exemplary models for high-density
matter in NSs~\citep{2000A&A...360..171R,2007A&A...467..395O,2011ApJ...738L..32G,2012arXiv1212.0905H} or even only
with a single NS EoS~\citep{2011ApJ...736L..21R,2012arXiv1204.6242P,2012arXiv1204.6240R,2012arXiv1206.2379K,2012arXiv1210.6549R},
although the importance of the nuclear EoS for a quantitative assessment
of the dynamical mass ejection can be concluded from published
calculations~\citep[e.g.][]{2011ApJ...738L..32G}. These calculations, however,
also suggest that the nuclear abundance pattern produced by r-processing
in the ejecta may be largely insensitive to variations of the
conditions in the ejecta.

It is important to note that quantitatively reliable information on the
ejecta masses and their dependence on the binary and EoS properties 
require general relativistic (GR) simulations. Newtonian results in the 
literature~\citep{1999A&A...341..499R,1999ApJ...527L..39J,2001A&A...380..544R,2011ApJ...736L..21R,2012arXiv1206.2379K,2012arXiv1204.6240R,2012arXiv1204.6242P,2012arXiv1210.6549R} exbibit significant quantitative and qualitative differences compared
to relativistic models~\citep{2007A&A...467..395O,2011ApJ...738L..32G,2012arXiv1212.0905H}. Newtonian calculations tend to overestimate the ejecta masses in general~\citep{1999A&A...341..499R,1999ApJ...527L..39J,2001A&A...380..544R,2011ApJ...736L..21R,2012arXiv1206.2379K,2012arXiv1204.6240R,2012arXiv1204.6242P,2012arXiv1210.6549R}. This can be understood because of several facts. First,
the structure of NSs in GR is considerably more compact than that of 
Newtonian stars. For instance, a NS with a gravitational mass
of 1.35\,$M_{\odot}$ described by the LS220 EoS~\citep{1991NuPhA.535..331L}
possesses a circumferential radius of 12.6\,km, whereas the corresponding 
Newtonian star has 14.5\,km. Second, GR gravity is stronger and the 
merging of two NSs is therefore more violent. The difference can be expressed
in terms of the gravitational binding energy of a nucleon on the 
surface of the considered 1.35\,$M_{\odot}$ NSs, which is $\sim$200\,MeV 
in the GR case compared to only $\sim$130\,MeV for the Newtonian model.
Third, GR forces merger remnants beyond a mass limit to collapse
to black holes on a dynamical timescale. Such an effect cannot be tracked
by Newtonian models. These differences are of direct relevance for the
collision dynamics and the possibility to unbind matter from the inner
and outer crust regions of the merging NSs.

It is the purpose of this paper to explore the influence of the 
high-density EoS on the ejecta properties in a systematic way, i.e.,
we will determine ejecta masses and the nucleosynthesis outcome for 
a large set of NS matter models, applying them in relativistic NS merger
simulations. Most of these EoSs were already
employed in our previous works~\citep{2012PhRvL.108a1101B,2012PhRvD..86f3001B}. They were chosen such that they provide as completely as 
possible a coverage of the possibilities for NS properties
(expressed by corresponding mass-radius-relations) which are compatible
with present observational constraints (e.g., the 1.97\,$M_\odot$ NS 
discovery of~\citet{2010Natur.467.1081D}, and more recently 2.01\,$M_\odot$ by~\citet{Antoniadis26042013}) and theoretical understanding~\citep{2010arXiv1012.3208L,2007PhR...442..109L,2010ApJ...722...33S,2010PhRvL.105p1102H}.
In our study we will focus on symmetric 1.35-1.35\,$M_{\odot}$ systems
and will compare them with asymmetric 1.2-1.5\,$M_{\odot}$ mergers. 
Because population synthesis models~\citep{2008ApJ...680L.129B} and
pulsar observations~\citep{1999ApJ...512..288T,2011A&A...527A..83Z} suggest
that the double NS population is strongly dominated by systems of 
nearly equal-mass stars of about 1.35\,$M_{\odot}$ each, the average NS merger
event can be well represented by a 1.35-1.35\,$M_{\odot}$ configuration,
and a clarification of the EoS dependence of ejecta masses, 
r-process yields, and properties of electromagnetic counterparts
of NS mergers seems to be more important than a wide variation of binary
parameters. Nevertheless, we will also present results of a more 
extended survey of binary mass ratios and total masses for some
representative EoSs.

In our work we will exclusively concentrate on NS-NS mergers, but
the discussed phenomena should play a role also for
NS-black hole coalescence~\citep{1999ApJ...527L..39J,2000MNRAS.318..606L,2004MNRAS.351.1121R,2005ApJ...634.1202R,2006PhRvD..73b4012F,2012arXiv1212.4810F,2012arXiv1210.6549R,2012arXiv1204.6242P,2012arXiv1204.6240R} and 
eccentric NS mergers~\citep{2012PhRvD..85l4009E,2012ApJ...760L...4E,2012arXiv1204.6240R,2012arXiv1210.6549R}. 
However, while the existence of double NS systems is established
by observations, progenitors of NS-black
hole and eccentric NS mergers have not been observed yet and
the rates of such types of events are even more uncertain
than those of coalescing binary NSs. In investigating the latter we
will only consider the phase of dynamical 
mass ejection between about the time when the two NSs collide until 
a few milliseconds later. During this phase hydrodynamical and 
tidal forces (shock compression, pressure forces, gravitational
interaction) are responsible for the mass shedding of the merging
objects. Once the remnant has formed, however, differential rotation
is expected to strongly amplify the magnetic fields~\citep[e.g.][]{2006Sci...312..719P,2008PhRvD..77b4006A,2008PhRvD..78b4012L,2011PhRvD..83d4014G} and viscous energy dissipation is likely
to provide additional heating, enhancing the neutrino emission that
accompanies the secular evolution of the post-merger configuration~\citep{1999A&A...344..573R,2004MNRAS.352..753S,2009ApJ...690.1681D,2012PThPh.127..535S,2013arXiv1304.6720F}. 
As a consequence the merger remnant will experience mass loss due to
neutrino energy deposition in the near-surface regions~\citep{1999A&A...344..573R,2004MNRAS.352..753S,2009ApJ...690.1681D,2012ApJ...746..180W,2012PThPh.127..535S} (similar to the 
neutrino-driven wind of proto-neutron stars emerging from stellar core
collapse) and due to magnetohydrodynamical outflows. Both mechanisms
will add ejecta to the mass stripped during the dynamical interaction 
of the system components, but the details of the secular evolution and 
the associated mass loss will be very sensitive to the 
EoS-dependent stability properties of the merger remnant, i.e.,
to the question whether the remnant is a hypermassive NS (see~\citet{2000ApJ...528L..29B} for a definition) or 
whether and when it collapses to a black hole-torus system. 
These questions lie beyond the scope of the present work.

Our paper is organized as follows. In Sect.~\ref{sec:code} a brief 
summary of the numerical methods and microphysics ingredients
of our NS merger simulations is given.
In Sect.~\ref{sec:masses} we present our results for the 
relation between dynamical mass loss and NS (EoS) properties,
provide a detailed description of the mass-loss dynamics in our
relativistic models (drawing comparisons to Newtonian results), 
discuss the influence of an approximate
treatment of thermal effects in the EoS, and evaluate the mass
ejection for three selected, representative EoSs in merger
simulations for a wider space of binary masses and mass ratios
in order to determine the population-integrated mass loss.
In Sect.~\ref{sec:nucleo} we describe results of nuclear network
calculations performed for a subset of our merger models and
draw conclusions on the Galactic merger rate and the production
of long-lived radioactive species ($^{232}$Th, $^{235}$U, $^{238}$U)
used for stellar nucleocosmochronometry. Finally, we present
values for the heating efficiency of the merger ejecta by 
radioactive decays of the nucleosynthesis products and apply
them in Sect.~\ref{sec:opttrans} to estimate the properties
(peak luminosity, peak timescale, effective temperature
at the maximum luminosity) of the optical transients that can be
expected from the expanding merger debris. We also briefly discuss
the implications of our simulations for radio flares. Finally,
a summary and conclusions follow in Sect~\ref{sec:sum}.

\section{Numerical model and equations of state} \label{sec:code}
The simulations of our study are performed with a relativistic Smooth Particle Hydrodynamics (SPH) code, i.e. the hydrodynamical equations are evolved in a Lagrangian manner~\citep{2002PhRvD..65j3005O,2007A&A...467..395O,2010PhRvD..82h4043B}. The Einstein field equations are solved imposing conformal flatness of the spatial metric~\citep{1980grg..conf...23I,1996PhRvD..54.1317W}, and a gravitational-wave backreaction scheme is used to account for energy and angular momentum losses by the emission of gravitational radiation~\citep{2007A&A...467..395O}. The code evolves the conserved rest-mass density $\rho^{*}$, the conserved specific momentum $\tilde{u}_i$, and the conserved energy density $\tau$, whose definitions evolve the metric potentials and the ``primitive'' hydrodynamical quantities, i.e. the rest-mass density $\rho$, the coordinate velocity $v_i$, and the specific internal energy $\epsilon$. The system of relativistic hydrodynamical equations is closed by an EoS, which relates the pressure $P=P(\rho,T,Y_{\mathrm{e}})$ and the specific internal energy $\epsilon=\epsilon(\rho,T,Y_{\mathrm{e}})$ to the rest-mass density $\rho$, the temperature $T$ and the electron fraction $Y_{\mathrm{e}}$. The temperature is obtained by inverting the specific internal energy $\epsilon=\epsilon(\rho,T,Y_{\mathrm{e}})$ for given $\rho$ and $Y_{\mathrm{e}}$. Changes of the electron fraction are assumed to be slow compared to the dynamics~\citep[see e.g.][]{1997A&A...319..122R}, and the initial electron fraction, which is defined by the neutrinoless beta-equilibrium of cold NSs, is advected according to $\frac{\mathrm{d} Y_{\mathrm{e}}}{\mathrm{d}t}=0$ ($\frac{d}{dt}$ defines the Lagrangian, i.e. comoving, time derivative).

The EoS of NS matter is only incompletely known and numerical studies rely on theoretical prescriptions of high-density matter. This work surveys a representative sample of 40 microphysical EoSs, which have been derived within different theoretical frameworks and make different assumptions about the composition of high-density matter and the description of nuclear interactions. Most of the employed EoSs are listed in~\citet{2012PhRvD..86f3001B}, where details can be found, while some new models are introduced below. Because of the one-to-one correspondence between the EoS and the mass-radius relation of nonrotating NSs, it is convenient to characterize EoSs by the resulting stellar properties. Stellar quantities as integral properties of an EoS are in particular useful to classify the dynamics of NS mergers and the accompanying gravitational-wave signals~\citep{2012PhRvL.108a1101B,2012PhRvD..86f3001B}. For this reason we will adopt the same approach also for this investigation. Considering for instance NSs with a gravitational mass of 1.35~$M_{\odot}$, the stellar radii $R_{1.35}$ vary from 10.13~km to 15.74~km for the different EoSs of our sample. The maximum mass $M_{\mathrm{max}}$ of nonrotating NSs obtained for these EoSs ranges from 1.79~$M_{\odot}$ to 3.00~$M_{\odot}$. In terms of their stellar properties the employed EoSs of our study show a large variation (see the mass-radius relations in Fig.~4 of~\citet{2012PhRvD..86f3001B}). Note that we do not apply any selection procedure for choosing the EoSs, except that we require a maximum mass above $\approx 1.8~M_{\odot}$. This limit is chosen because of the firm discovery of pulsars with gravitational masses of $(1.97\pm 0.04)~M_{\odot}$~\citep{2010Natur.467.1081D} and $(2.01\pm 0.04)~M_{\odot}$~\citep{Antoniadis26042013}. EoSs which yield a maximum mass below this limit are practically excluded by this observation. Nevertheless, we accept them (at least down to $M_{\mathrm{max}}\approx 1.8~M_{\odot}$) for our investigation because we expect that at densities relevant in a typical NS merger these models still provide a viable description of high-density matter~\citep[see][]{2012PhRvD..86f3001B}. Note that compared to our previous study in~\citet{2012PhRvD..86f3001B} we extend our EoS survey by including also the models TM1, TMA, NL3, DD2, SFHO and SFHX of~\citet{2010NuPhA.837..210H}~\citet{2012ApJ...748...70H} and~\citet{2012arXiv1207.2184S}, relying on the interactions described in~\citet{1994NuPhA.579..557S}~\citet{1995NuPhA.588..357T},~\citet{1997PhRvC..55..540L},~\citet{2010PhRvC..81a5803T} and~\citet{2012arXiv1207.2184S}. Moreover, we include the BSk20 and BSk21 EoSs of~\citet{2010PhRvC..82c5804G}. The maximum masses $M_{\mathrm{max}}$ resulting for these EoSs are 2.21~$M_{\odot}$, 2.02~$M_{\odot}$, 2.79~$M_{\odot}$, 2.42~$M_{\odot}$, 2.06~$M_{\odot}$, 2.13~$M_{\odot}$, 2.16~$M_{\odot}$ and 2.28~$M_{\odot}$ (order as listed above), while the radii of cold 1.35~$M_{\odot}$ NSs are 14.49~km, 13.86~km, 14.75~km, 13.21~km, 11.92~km, 11.98~km, 11.74~km and 12.54~km, respectively. From our sample of EoSs in~\citet{2012PhRvD..86f3001B} we do not consider the SKA EoS (because of its restriction to densities above $1.7\times 10^9~\mathrm{g/cm^3}$) and EoSs which are not compatible with the pulsar observation of~\citet{2010Natur.467.1081D,Antoniadis26042013} and directly form a black hole after merging. We also exclude absolutely stable strange quark matter. We refer to~\citet{2009PhRvL.103a1101B} for the particular implications of ejecta from strange quark star mergers.

Only 12 out of the considered 40 EoSs describe thermal effects consistently and provide the dependence of thermodynamical quantities on the temperature and the electron fraction. Instead, the majority of models considers matter at zero temperature and in equilibrium with respect to weak interactions (i.e. for beta-equilibrium for neutrino-less conditions). Because temperature effects become important during the merging of the binary components and during the subsequent evolution, we employ an approximate treatment of thermal effects for those EoSs which are given as barotropic relations. This procedure supplements the pressure by an additional ideal-gas component to mimic thermal pressure support, and it requires to choose a corresponding ideal-gas index $\Gamma_{\mathrm{th}}$. Appropriate values for $\Gamma_{\mathrm{th}}$ are in the range of 1.5 to 2 for high-density matter~\citep{2010PhRvD..82h4043B}. The uncertainties connected to the use of this approximate temperature description and the choice of the ideal-gas index were examined in~\citet{2010PhRvD..82h4043B}, where also details about the exact implementation can be found.

From population synthesis studies~\citep{2008ApJ...680L.129B} and in agreement with pulsar observations~\citep{1999ApJ...512..288T,2011A&A...527A..83Z} it is expected that binaries with two NSs with gravitational masses of about $M_1\approx M_2\approx 1.35~M_{\odot}$ are the most abundant systems. For this reason we focus in our EoS survey on such equal-mass binaries, albeit we also explore the influence of a system asymmetry by considering 1.2-1.5~$M_{\odot}$ binaries. For a selected subset of EoS models the full range of possible binary parameters is investigated, varying the single component masses from 1.2~$M_{\odot}$ to approximately the maximum mass of NSs.

Because of energy and angular momentum losses by gravitational radiation the orbits of NS binaries shrink and the binary components merge after an inspiral period, which lasts roughly 100 to 1000 Myrs for the known systems~\citep{2008LRR....11....8L}. The typical outcome of the coalescence of a 1.35-1.35~$M_{\odot}$ binary system is the formation of a differentially rotating object, potentially a hypermassive NS (i.e. a NS that is more massive than the maximum-mass rigid-rotation configuration and that is stabilized temporarily by differential rotation~\citep{2000ApJ...528L..29B}). The merger remnant is surrounded by an extended halo structure of low-density material. Only four EoSs of our sample lead to the prompt formation of a black hole within about one millisecond after the collision because the remnant can not be supported against the gravitational collapse. For a description of the general dynamics and a more thorough discussion of the collapse behavior we refer to~\citet{2007A&A...467..395O},~\citet{2010PhRvD..82h4043B} and~\citet{2012PhRvD..86f3001B}. In this paper only initially nonrotating NSs are investigated because viscosity is too low to yield tidally locked systems. The stars in NS binaries are therefore expected to rotate slowly in comparison to the orbital angular velocity, justifying the use of an irrotational velocity profile~\citep{1992ApJ...400..175B,1992ApJ...398..234K}.

In this study we analyze the material which becomes gravitationally unbound during or right after merging. In order to estimate whether a given fluid element, i.e. an SPH particle, can escape to infinity, we consider
\begin{equation}\label{eq:ejrel}
\epsilon_{\mathrm{stationary}}=v^i\tilde{u}_i+\frac{\epsilon}{u^0}+\frac{1}{u^0}-1>0
\end{equation}
with the coordinate velocity $v^i$, the conserved momentum $\tilde{u}_i$, and the time-component of the eigen-velocity $u^0$ (in geometrical units). This expression can be derived from the hydrodynamical equations by neglecting pressure forces and assuming a stationary metric~\citep{2002PhRvD..65j3005O}. The quantity $\epsilon_{\mathrm{stationary}}$ is conserved $\left( \frac{d \epsilon_{\mathrm{stationary}}}{dt}=0 \right)$ and at infinity it reduces to the Newtonian expression for the total energy of a fluid element. Hence, a particle with  $\epsilon_{\mathrm{stationary}}>0$ will be unbound. Equation~\eqref{eq:ejrel} is evaluated in a time-dependent way and SPH particles that fulfill this criterion 10~ms after merging are considered as ultimately gravitationally unbound. Note that our simulations neglect a possible (smaller) contribution to the ejecta by neutrino-driven winds or magnetically driven outflows from the secular evolution of the merger remnant~\citep{2009ApJ...690.1681D,2012ApJ...746..180W}.

\section{Ejecta masses}\label{sec:masses}
\begin{figure}
\includegraphics[width=8.6cm]{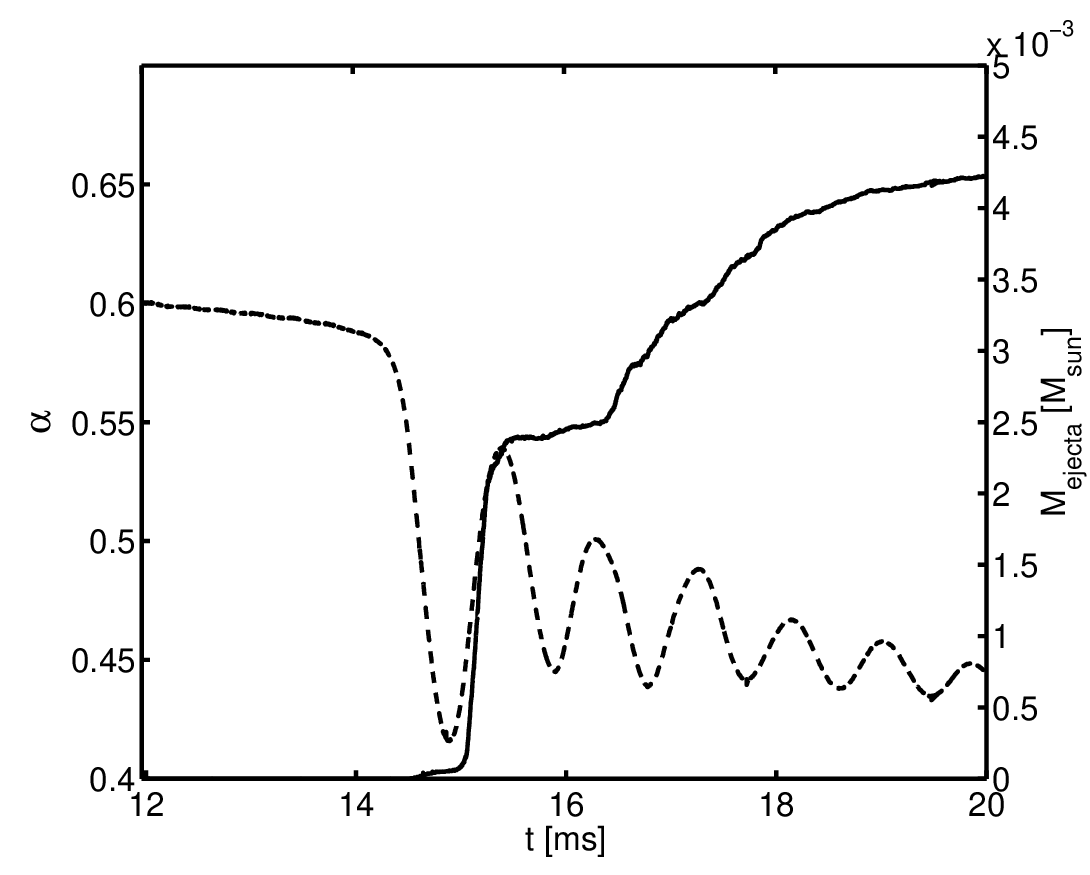}
\includegraphics[width=8.6cm]{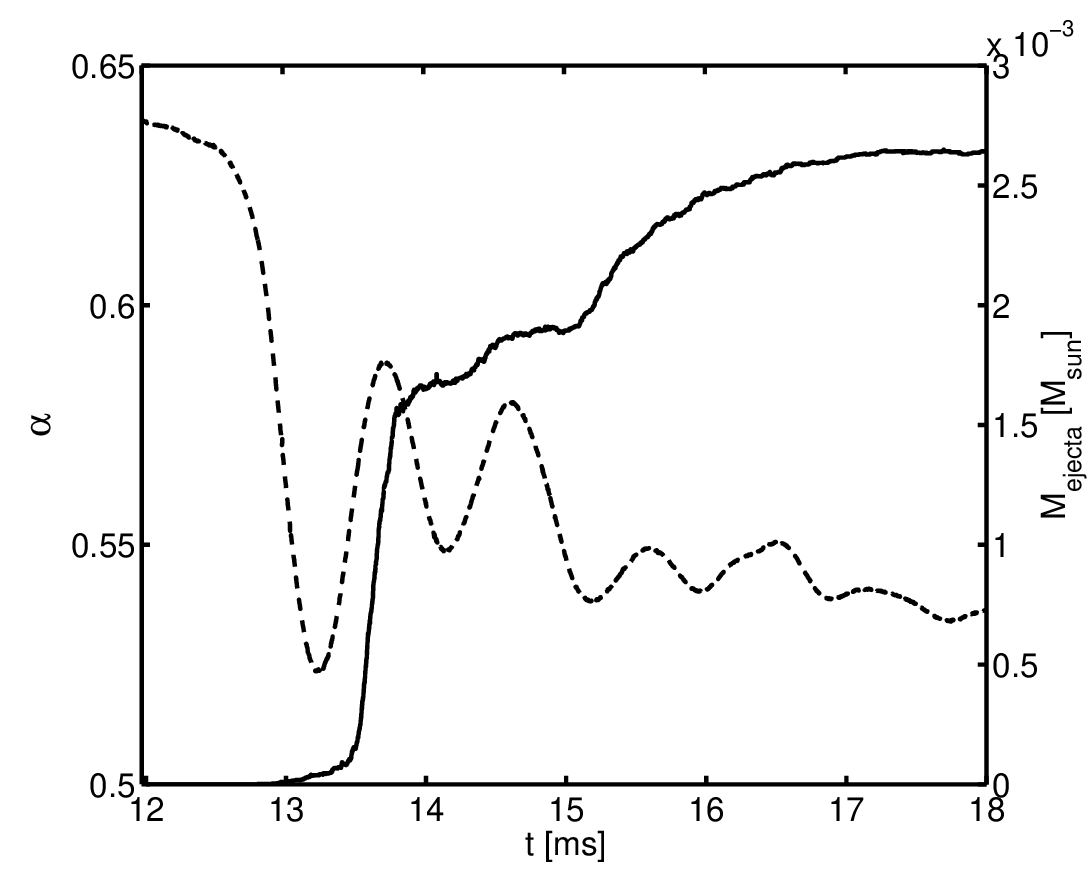}
\includegraphics[width=8.6cm]{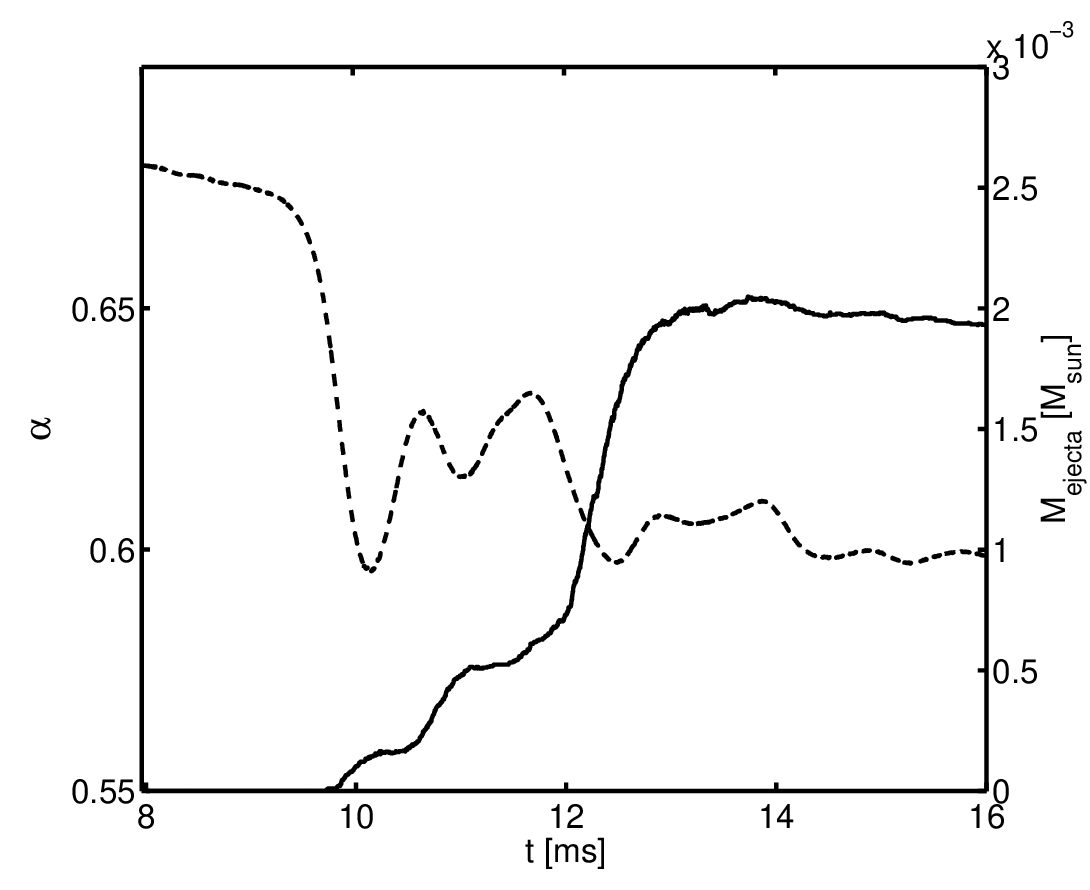}
\caption{\label{fig:lapse}Evolution of the minimum lapse function $\alpha$ (dashed line) and the amount of unbound matter (solid line) for the symmetric 1.35-1.35~$M_{\odot}$ merger with the soft SFHO EoS (top panel), the intermediate DD2 EoS (middle panel), and the stiff NL3 EoS (bottom panel).}
\end{figure}

In the following we employ Eq.~\eqref{eq:ejrel} to determine the unbound material in different merger simulations. After a first steep rise of the ejecta mass shortly after the merging of the two NSs, the mass fulfilling the ejecta criterion remains approximately constant (Fig.~\ref{fig:lapse}). A few models, however, show a continuing, slow increase of the ejecta mass also at later times. The ejecta masses discussed below are computed 10~ms after merging.

In order to determine the influence of the high-density EoS on the ejecta we employ approximately the same numerical resolution of about 350,000 SPH particles for all simulations. By using nonuniform SPH particle masses to model the stellar profile (more massive particles in the high-density core and lighter particles in the outer low-density layers) it is possible to achieve a better resolution of low-density regions. This results in an effective mass resolution of about $2\times 10^{-6}~M_{\odot}$, which is comparable to a SPH simulation of about one million equal-mass particles. The influence of the numerical resolution is investigated by performing additional simulations with higher SPH particle numbers. For the TM1 EoS the ejecta masses are found to be (in solar masses) $1.67\times 10^{-3}, 1.80\times 10^{-3}, 1.71\times 10^{-3}, 2.43\times 10^{-3}$, and $2.07\times 10^{-3}$ for calculations with about $339\times 10^3, 550\times 10^3, 782\times 10^3, 1007\times 10^3$, and $1272\times 10^3$ SPH particles. In these simulations 521, 838, 1241, 2275, and 3080 particles are ejected. Determining the unbound matter 6~ms after merging for the APR EoS, we find ejecta masses (in solar masses) of $5.93\times 10^{-3}, 6.08\times 10^{-3}, 6.54\times 10^{-3}, 6.69\times 10^{-3}$, and $6.14\times 10^{-3}$ in simulations with $339\times 10^3, 592\times 10^3, 782\times 10^3, 1007\times 10^3$, and $1272\times 10^3$ SPH particles (with 2488, 4473, 6461, 9168, and 16958 particles unbound). Thus, the numerical resolution has an effect on the level of some 10 per cent. This, however, is smaller than the impact of the EoS (see below), which is the focus of this paper. The nonmonotonic variations of the ejecta mass with increasing resolution indicate that statistical fluctuations have some influence on the ejected particle population as well. Note that a small fraction of weakly bound matter could become unbound by the heat generated in the nucleosynthesis processes~\citep{2010MNRAS.402.2771M}, which is not taken into account in our hydrodynamical simulations.

\subsection{Origin of the ejecta and comparison with other calculations}
\begin{figure*}
\includegraphics[width=8.9cm]{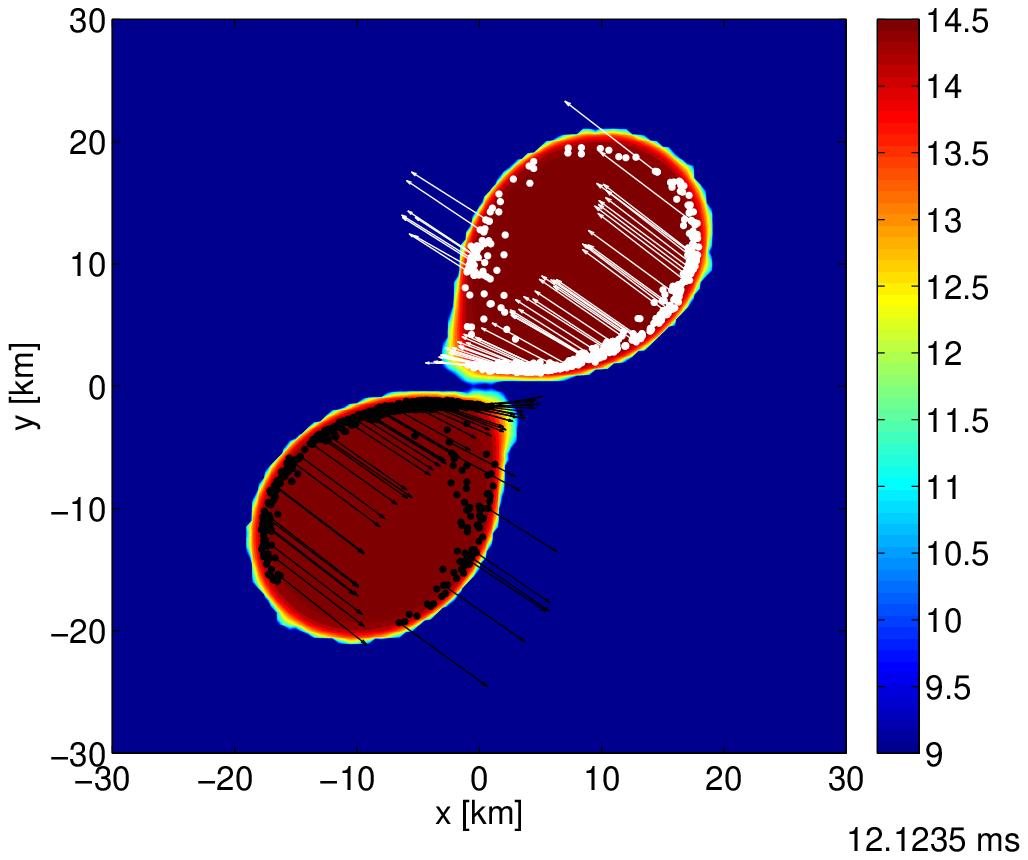}
\includegraphics[width=8.9cm]{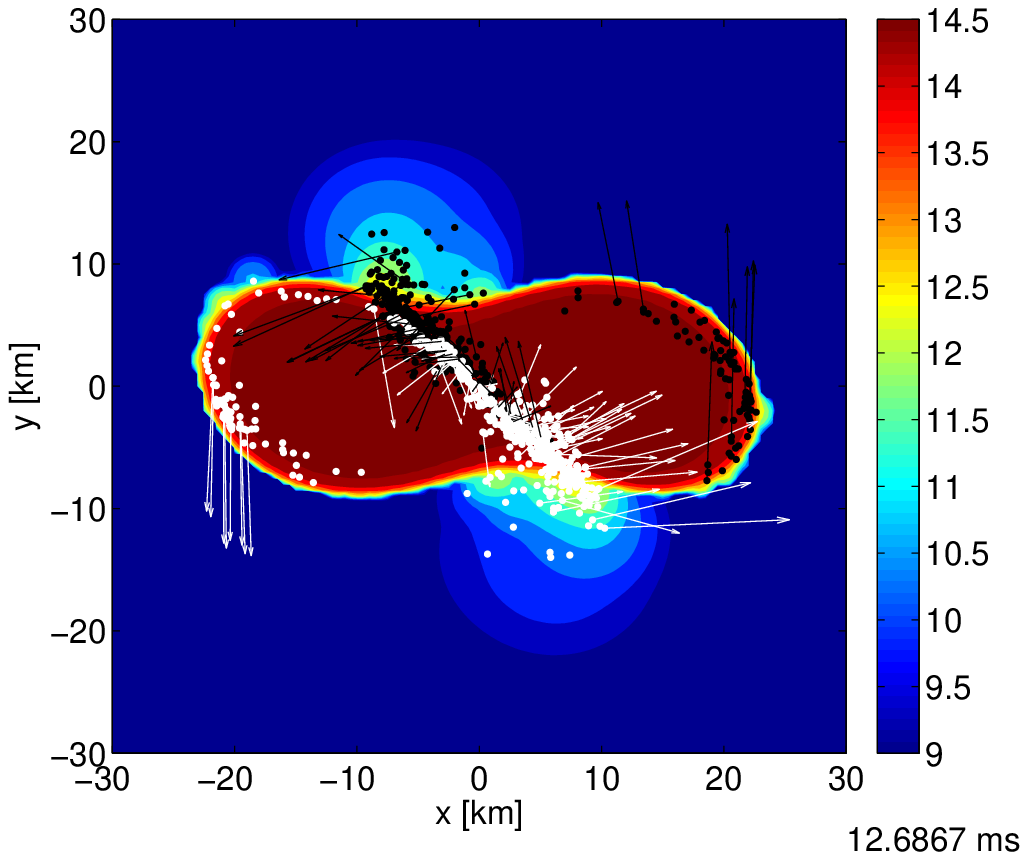}

\includegraphics[width=8.9cm]{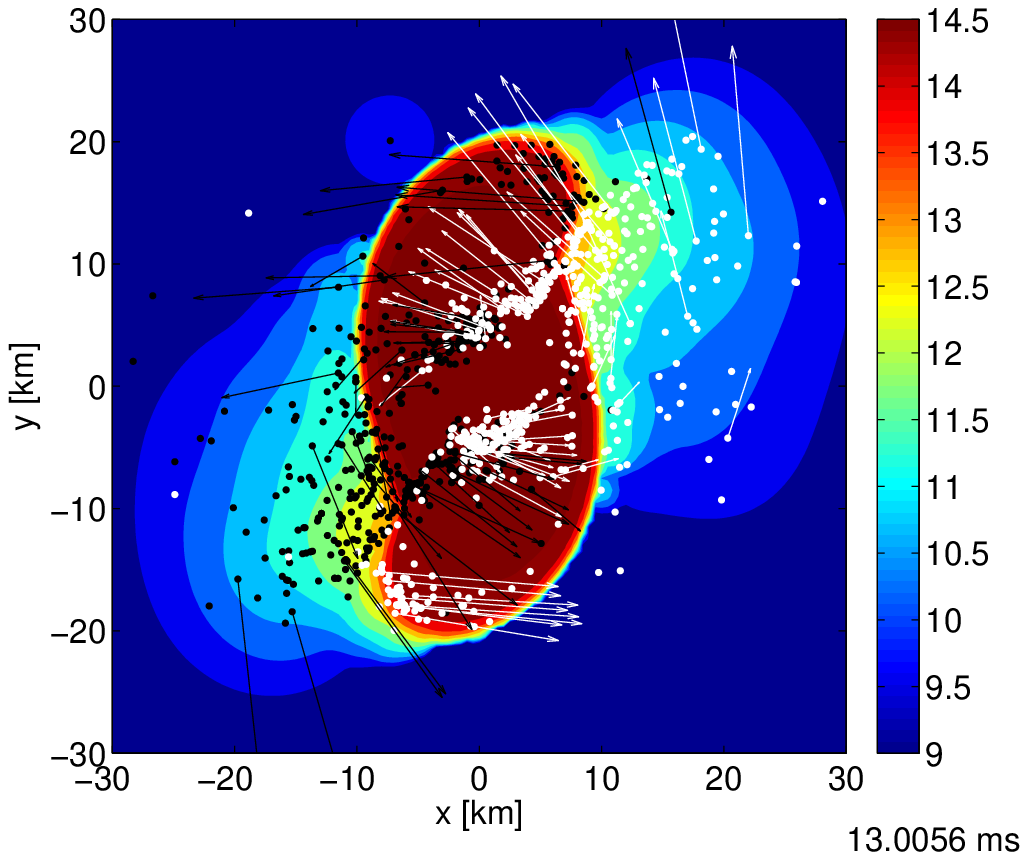}
\includegraphics[width=8.9cm]{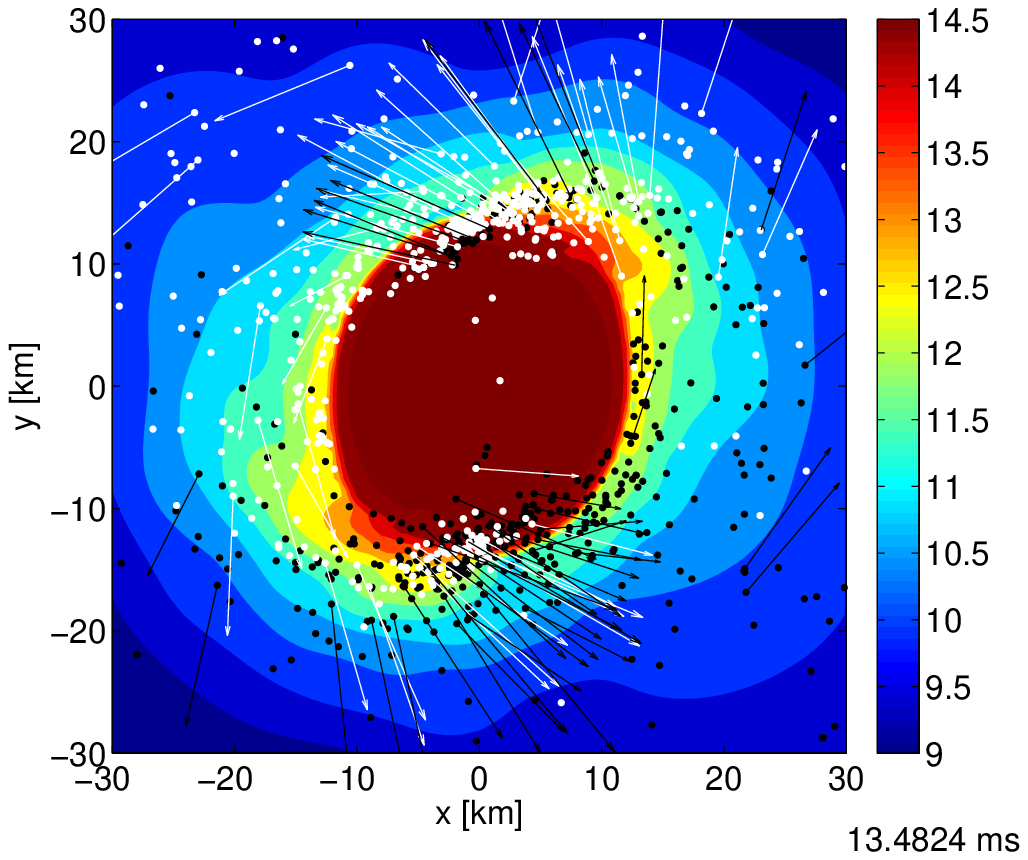}

\includegraphics[width=8.9cm]{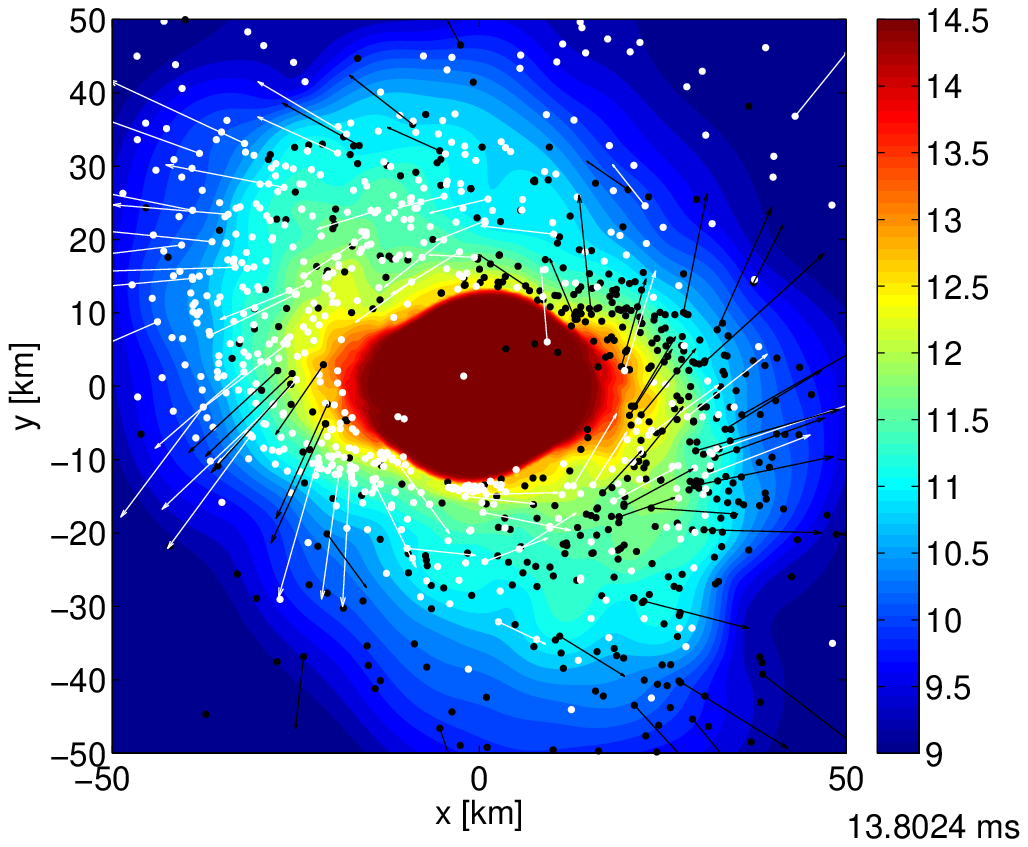}
\includegraphics[width=8.9cm]{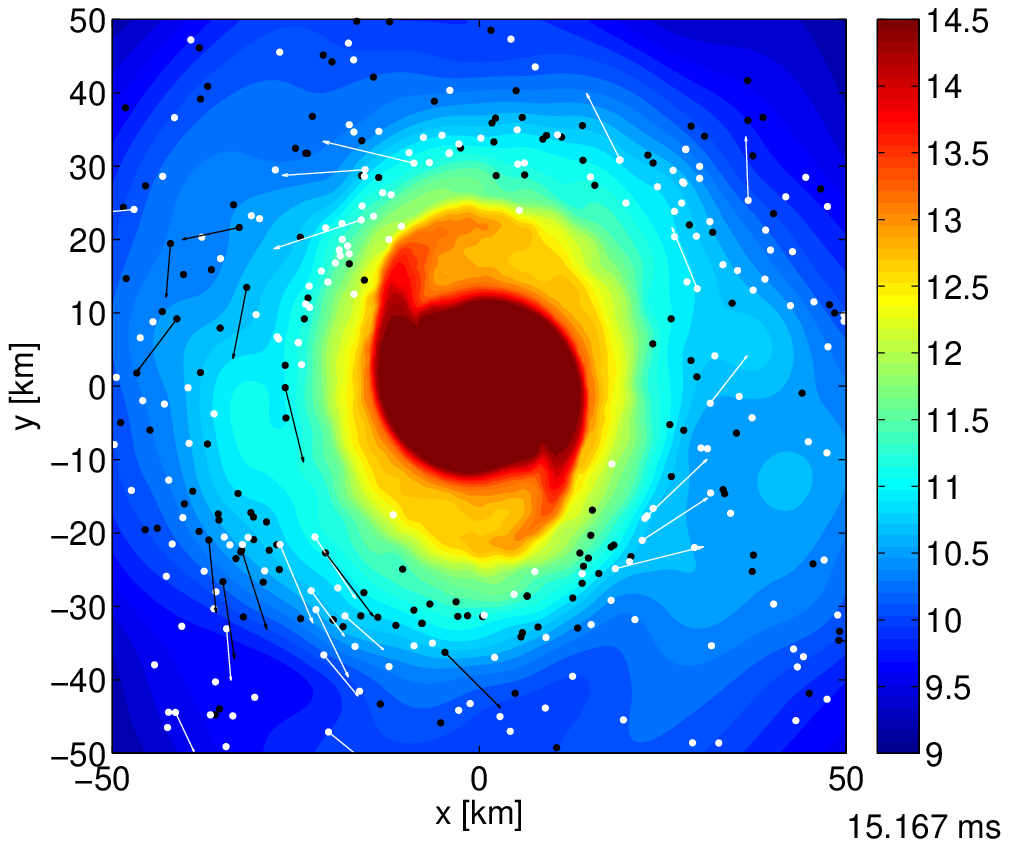}
\caption{\label{fig:snap}Merger and mass ejection dynamics of the 1.35-1.35~$M_{\odot}$ binary with the DD2 EoS, visualized by the color-coded conserved rest-mass density (logarithmically plotted in $\mathrm{g/cm^3}$) in the equatorial plane. The dots mark SPH particles which represent ultimately gravitationally unbound matter. Their positions are projections of the three-dimensional locations anywhere in the merging stars onto the orbital plane. Black and white indicate the origin from the one or the other NS. For every tenth particle the coordinate velocity is indicated by an arrow with a length proportional to the absolute value of the velocity (the speed of light corresponds to a line length of 50~km). The time is indicated below the color bar of each panel. Note that the side length of the bottom panels is enlarged.  The visualization tool SPLASH was used to convert SPH data to grid data~\citep{2007PASA...24..159P}.}
\end{figure*}

As can be seen in Fig.~\ref{fig:snap} most of the ejecta originate from the contact interface between the colliding binary components, which get deformed into drop-like shapes prior to the merging. For the 1.35-1.35~$M_{\odot}$ binary the ejecta in the shear interface between the stars are separate into two components, each being fed (nearly) symmetrically by material from both colliding stars (top right panel and middle left panel). The matter in the cusps of the stars essentially keeps its direction of motion towards the companion, whereas the backward part of the contact interface mixes with some of the companion matter (top right panel). Both lumps of ejecta are squeezed out from the contact interface and expand on the retral side of the respective companion star, partially slipping over it (middle panels). The bulk matter of the binary components forms a rotating double-core structure where the two dense cores oscillate against each other (not visible because of the logarithmic density scale; see e.g. the descriptions in~\citet{2010PhRvD..81b4012B} and~\citet{2011MNRAS.418..427S}). A first bunch of the matter which was squeezed out from the contact sheet gets unbound in a first expansion phase of the rotating double core structure, which pushes the ejecta outward. This can be seen in Fig.~\ref{fig:snap} (middle right panel and bottom left panel) and also in the evolution of the minimum lapse function $\alpha$ (Fig.~\ref{fig:lapse}), which is a measure for the compactness of the central object. As the cores separate from each other and the lapse function grows out of its minimum, the ejecta mass increases. A second expansion of the double cores unbinds a smaller amount of matter. Finally, about two milliseconds after the first contact, the triaxial deformation grows into two spiral-arm like extensions reaching out from the central remnant. These expand into the surrounding, low-density halo fed from the contact interface, push it away and unbind additional matter (bottom right panel of Fig.~\ref{fig:snap} and second mass-loss episode visible in Fig.~\ref{fig:lapse}). For different EoSs the different dynamical mechanisms contribute to the ejecta production with different relative strengths. For soft EoSs (top panel of Fig.~\ref{fig:lapse}) the first steep rise of the ejecta mass due to the expanding double core is much more pronounced, whereas for stiff EoSs (bottom panel of Fig.~\ref{fig:lapse}) the first increase of the ejecta is very moderate and the late spiral arms unbind most of the ejecta. This can be seen in Fig.~\ref{fig:lapse} for the SFHO EoS representing a soft EoS, for the NL3 EoS as a stiff example, and the intermediate case of the DD2 EoS.

A significantly smaller fraction of the ejecta (typically below 25 per cent) stems from the outer faces of the merging stars opposite to the contact layer (SPH particles at the outer left and outer right ends of the stellar body in the top right panel of Fig.~\ref{fig:snap}). During merging this matter at the rear of the star (SPH particles with nearly horizontal velocity vectors at the top and bottom merger tails in the middle left panel) lags behind the rotation of the star's center and is hit and ablated by the ``nose'' of the companion shortly after the snapshot shown in the middle left panel (see velocity arrows of particles with opposite color at the tips of the noses). This material gets mostly unbound in the first expansion phase of the oscillating double-core structure. Such type of ejecta is less abundant for stiff EoSs.

Different from relativistic simulations, Newtonian models find the ejecta originating mostly from the tips of tidal tails~\citep[see e.g.][]{2012arXiv1206.2379K}, in particular also in the case of symmetric binaries. Relativistic calculations (within the CFC framework)~\citep{2007A&A...467..395O,2011ApJ...738L..32G} yield the dominant ejection from the contact interface as described above (see also the inset of Fig.~1 in~\citet{2011ApJ...738L..32G}). Recently, the fully relativistic simulations of~\citet{2012arXiv1212.0905H} have provided further support for the ejecta origin from the contact interface, confirming the conclusions of~\citet{2007A&A...467..395O} and~\citet{2011ApJ...738L..32G}. This points to qualitative differences between the Newtonian and relativistic mass-loss dynamics with the important difference that in relativistic simulations all of the ejecta are shock-heated while in Newtonian calculations the cold, tidally stripped material dominates.

A quantitative comparison between Newtonian~\citep{2012arXiv1206.2379K,2012arXiv1204.6240R,2012arXiv1204.6242P,2012arXiv1210.6549R} and relativistic simulations~\citep{2007A&A...467..395O,2011ApJ...738L..32G} also reveals considerable discrepancies. For instance, simulations of a 1.4-1.4~$M_{\odot}$ merger with the Shen EoS in Newtonian theory produce more than $10^{-2}~M_{\odot}$ ejecta~\citep{2012arXiv1206.2379K,2012arXiv1204.6240R,2012arXiv1204.6242P,2012arXiv1210.6549R}, whereas the relativistic calculations of this study and in~\citet{2007A&A...467..395O} and~\citet{2011ApJ...738L..32G} yield only a few $10^{-3}~M_{\odot}$ of unbound material for the 1.35-1.35~$M_{\odot}$ binary with the same EoS. Comparing the results of our study with the likewise relativistic calculations in~\citet{2012arXiv1212.0905H} shows very good agreement for all four EoSs used in~\citet{2012arXiv1212.0905H}. For example, for the APR EoS with $\Gamma_{\mathrm{th}}=2$ both groups find about $5\times 10^{-3}~M_{\odot}$ of unbound matter. This is remarkable because the implementations differ with respect to the hydrodynamics scheme, which is an SPH (smooth particle hydrodynamics) algorithm here but a grid-based, high-resolution central scheme in~\citet{2012arXiv1212.0905H}. (Note that we employ the conformal flatness approximation whereas the calculations in~\citet{2012arXiv1212.0905H} are conducted within full general relativity.) These findings provide confidence in the results on the quantitative level and point towards fundamental differences between Newtonian and relativistic treatments. Such differences are not unexpected because NSs are more compact in general relativity than in Newtonian gravity. The stronger gravitational attraction prevents the formation of pronounced tidal tails at the outer faces of the colliding stars and increases the strength of the collision.

\subsection{Equation of state dependence}
Several NS EoSs have been employed in merger simulations by different groups, but a large, systematic investigation of the EoS dependence of the ejecta production is still missing in particular with a consistent description of thermal effects. For a given EoS the radius $R_{1.35}$ of a nonrotating NS with 1.35~$M_{\odot}$ is a characteristic quantity specifying the compactness of NSs. Therefore, we use $R_{1.35}$ to describe the influence of the high-density EoS on the amount of NS merger ejecta.

\begin{table}
\begin{ruledtabular}
\caption{\label{tab:models}Model properties}
\begin{tabular}{l c c c c c}


\tableline \tableline
Model $M_1$-$M_2$ & $\Gamma_{\mathrm{th}}$& $R_{1.35}$ & $M_{\mathrm{ej}}$ & $v$ & $E_{\mathrm{kin}}$ \\
& & (km) &$(10^{-3}~M_{\odot})$ & (c) & $(10^{50}$ erg) \\
\tableline
NL3 1.35-1.35        & full & 14.75 &   2.09  &   0.18  &   0.98 \\
NL3 1.35-1.35        & 2    & 14.75 &   1.57  &   0.34  &   2.03 \\
NL3 1.35-1.35        & 1.8  & 14.75 &   1.60  &   0.32  &   2.99 \\
NL3 1.35-1.35        & 1.5  & 14.75 &   1.86  &   0.30  &   1.98 \\
GS1 1.35-1.35        & full & 14.72 &   2.19  &   0.32  &   3.43 \\
Shen 1.35-1.35       & full & 14.56 &   2.33  &   0.23  &   4.80 \\
TM1 1.35-1.35        & full & 14.49 &   1.67  &   0.16  &   0.74 \\
TM1 1.35-1.35        & 2    & 14.49 &   1.37  &   0.36  &   2.02 \\
TM1 1.35-1.35        & 1.8  & 14.49 &   1.33  &   0.34  &   1.77 \\
TM1 1.35-1.35        & 1.5  & 14.49 &   1.53  &   0.32  &   1.86 \\
TMA 1.35-1.35        & full & 13.86 &   2.05  &   0.18  &   1.19 \\
LS375 1.35-1.35      & full & 13.56 &   2.58  &   0.30  &   3.39 \\
GS2 1.35-1.35        & full & 13.38 &   2.74  &   0.19  &   2.16 \\
DD2 1.35-1.35        & full & 13.21 &   3.07  &   0.22  &   2.18 \\
DD2 1.35-1.35        & 2    & 13.21 &   2.57  &   0.34  &   3.31 \\
DD2 1.35-1.35        & 1.8  & 13.21 &   2.26  &   0.32  &   2.61 \\
DD2 1.35-1.35        & 1.5  & 13.21 &   2.72  &   0.30  &   2.90 \\
LS220 1.35-1.35      & full & 12.64 &   1.99  &   0.28  &   2.08 \\
LS180 1.35-1.35 x  & full & 12.14 &   2.26  &   0.29  &   3.02 \\
SFHX 1.35-1.35       & full & 11.98 &   6.16  &   0.22  &   4.36 \\
SFHO 1.35-1.35       & full & 11.92 &   4.83  &   0.23  &   3.61 \\
SFHO 1.35-1.35       & 2    & 11.92 &   2.96  &   0.32  &   3.37 \\
SFHO 1.35-1.35       & 1.8  & 11.92 &   3.26  &   0.34  &   4.18 \\
SFHO 1.35-1.35       & 1.5  & 11.92 &   3.82  &   0.30  &   4.14 \\
eosL 1.35-1.35       & 2    & 15.74 &   1.49  &   0.22  &   0.77 \\
eosL 1.35-1.35       & 1.5  & 15.74 &   3.45  &   0.20  &   1.49 \\
MS1 1.35-1.35        & 2    & 14.99 &   1.17  &   0.27  &   0.98 \\
MS1 1.35-1.35        & 1.5  & 14.99 &   2.38  &   0.21  &   1.19 \\
MS1b 1.35-1.35       & 2    & 14.59 &   1.67  &   0.25  &   1.26 \\
MS1b 1.35-1.35       & 1.5  & 14.59 &   3.64  &   0.21  &   1.85 \\
Glenh3 1.35-1.35 x & 2    & 14.52 &   1.08  &   0.23  &   0.62 \\
Glenh3 1.35-1.35 x & 1.5  & 14.52 &   1.69  &   0.22  &   0.90 \\
MS2 1.35-1.35 x    & 2    & 14.25 &   0.81  &   0.26  &   0.65 \\
H3 1.35-1.35 x     & 2    & 13.95 &   1.43  &   0.27  &   1.15 \\
H4 1.35-1.35         & 2    & 13.95 &   1.28  &   0.27  &   1.09 \\
H4 1.35-1.35         & 1.5  & 13.95 &   1.93  &   0.27  &   1.64 \\
Heb6 1.35-1.35       & 2    & 13.33 &   1.55  &   0.45  &   3.86 \\
Heb6 1.35-1.35       & 1.5  & 13.33 &   3.43  &   0.24  &   2.58 \\
eosO 1.35-1.35       & 2    & 12.85 &   3.52  &   0.28  &   3.04 \\
eosO 1.35-1.35       & 1.5  & 12.85 &   4.62  &   0.25  &   3.27 \\
ALF2 1.35-1.35       & 2    & 12.78 &   3.80  &   0.28  &   3.36 \\
ALF2 1.35-1.35       & 1.5  & 12.78 &   4.49  &   0.27  &   3.80 \\
BSk21 1.35-1.35      & 2    & 12.54 &   3.36  &   0.32  &   3.89 \\
BSk21 1.35-1.35      & 1.5  & 12.54 &   4.37  &   0.27  &   3.81 \\
Heb4 1.35-1.35       & 2    & 12.51 &   1.89  &   0.43  &   4.33 \\
Heb4 1.35-1.35       & 1.5  & 12.51 &   2.41  &   0.39  &   5.13 \\
MPA1 1.35-1.35       & 2    & 12.49 &   3.64  &   0.30  &   3.60 \\
MPA1 1.35-1.35       & 1.5  & 12.49 &   4.48  &   0.29  &   4.35 \\
Heb5 1.35-1.35       & 2    & 12.38 &   2.63  &   0.43  &   5.93 \\
Heb5 1.35-1.35       & 1.5  & 12.38 &   2.90  &   0.38  &   5.89 \\
eosC 1.35-1.35 x   & 1.5  & 12.06 &   3.09  &   0.27  &   2.49 \\
ENG 1.35-1.35        & 2    & 12.05 &   5.29  &   0.29  &   5.01 \\
ENG 1.35-1.35        & 1.5  & 12.05 &   6.32  &   0.26  &   5.30 \\
APR3 1.35-1.35       & 2    & 12.04 &   4.65  &   0.30  &   4.69 \\
APR3 1.35-1.35       & 1.5  & 12.04 &   6.15  &   0.27  &   5.50 \\
Heb3 1.35-1.35       & 2    & 12.03 &   2.99  &   0.41  &   6.43 \\
Heb3 1.35-1.35       & 1.5  & 12.03 &   3.70  &   0.37  &   6.93 \\
BurgioNN 1.35-1.35   & 2    & 11.99 &   2.47  &   0.39  &   4.73 \\
BurgioNN 1.35-1.35   & 1.5  & 11.99 &   2.44  &   0.37  &   4.67 \\
SLy4 1.35-1.35       & 2    & 11.79 &   3.99  &   0.29  &   3.75 \\
SLy4 1.35-1.35       & 1.5  & 11.79 &   6.40  &   0.27  &   5.53 \\
BSk20 1.35-1.35      & 2    & 11.74 &   4.68  &   0.31  &   4.90 \\
BSk20 1.35-1.35      & 1.5  & 11.74 &   7.83  &   0.26  &   6.80 \\

\tableline
 \end{tabular}

 \end{ruledtabular} 
\end{table}
\setcounter{table}{0}
\begin{table}
 \begin{ruledtabular}
\caption{(Continued)}
\begin{tabular}{l c c c c c}
\tableline \tableline
Model $M_1$-$M_2$ & $\Gamma_{\mathrm{th}}$& $R_{1.35}$ & $M_{\mathrm{ej}}$ & $v$ & $E_{\mathrm{kin}}$ \\
& & (km) &$(10^{-3}~M_{\odot})$ & (c) & $(10^{50}$ erg) \\
\tableline

ALF4 1.35-1.35 x   & 2    & 11.60 &   5.70  &   0.30  &   6.07 \\
ALF4 1.35-1.35 x   & 1.5  & 11.60 &   7.40  &   0.29  &   7.65 \\
Heb2 1.35-1.35       & 2    & 11.42 &   4.95  &   0.34  &   7.90 \\
Heb2 1.35-1.35       & 1.5  & 11.42 &   5.01  &   0.37  &   9.07 \\
APR 1.35-1.35        & 2    & 11.33 &   5.96  &   0.31  &   6.37 \\
APR 1.35-1.35        & 1.5  & 11.33 &   7.38  &   0.30  &   7.90 \\
BB2 1.35-1.35 x    & 2    & 11.30 &   4.95  &   0.28  &   4.17 \\
eosUU 1.35-1.35      & 2    & 11.18 &   7.02  &   0.32  &   7.80 \\
eosUU 1.35-1.35      & 1.5  & 11.18 &   9.42  &   0.31  &   10.0 \\
Heb1 1.35-1.35       & 2    & 10.81 &   6.85  &   0.31  &   8.54 \\
Heb1 1.35-1.35       & 1.5  & 10.81 &   10.88 &   0.32  &   12.54 \\
eosAU 1.35-1.35      & 2    & 10.44 &   4.05  &   0.36  &   5.64 \\
eosAU 1.35-1.35      & 1.5  & 10.44 &   1.51  &   0.29  &   1.69 \\

NL3 1.2-1.5          & full & 14.75 &   7.95  &   0.19  &   4.50  \\
GS1 1.2-1.5          & full & 14.72 &   6.43  &   0.22  &   6.55  \\
Shen 1.2-1.5         & full & 14.56 &   5.66  &   0.30  &   8.33  \\
TM1 1.2-1.5          & full & 14.49 &   8.66  &   0.17  &   3.94  \\
TMA 1.2-1.5          & full & 13.86 &   10.21 &   0.20  &   6.40  \\
LS375 1.2-1.5        & full & 13.56 &   6.66  &   0.27  &   7.60  \\
GS2 1.2-1.5          & full & 13.38 &   10.69 &   0.18  &   6.14  \\
DD2 1.2-1.5          & full & 13.21 &   8.79  &   0.20  &   4.97  \\
LS220 1.2-1.5        & full & 12.64 &   13.22 &   0.18  &   7.34  \\
LS180 1.2-1.5 x    & full & 12.14 &   18.58 &   0.20  &   12.13 \\
SFHX 1.2-1.5         & full & 11.98 &   14.67 &   0.19  &   7.91  \\
SFHO 1.2-1.5         & full & 11.92 &   13.39 &   0.22  &   8.94  \\
NL3 1.2-1.2          & full & 14.75 &   2.15  &   0.17  &   0.91  \\
NL3 1.2-1.35         & full & 14.75 &   4.25  &   0.21  &   2.74  \\
NL3 1.2-1.6          & full & 14.75 &   9.96  &   0.19  &   5.57  \\
NL3 1.2-1.8          & full & 14.75 &  15.68  &   0.15  &   5.75  \\
NL3 1.35-1.5         & full & 14.75 &   2.72  &   0.24  &   2.25  \\
NL3 1.35-1.8         & full & 14.75 &  18.81  &   0.21  &  11.31  \\
NL3 1.5-1.5          & full & 14.75 &   1.70  &   0.20  &   1.04  \\
NL3 1.5-1.8          & full & 14.75 &   8.10  &   0.21  &   4.94  \\
NL3 1.6-1.6          & full & 14.75 &   3.74  &   0.22  &   2.59  \\
NL3 1.8-1.8          & full & 14.75 &   9.08  &   0.24  &   7.25  \\
NL3 1.35-2.0         & full & 14.75 &  12.85  &   0.20  &   7.62  \\
NL3 2.0-2.0          & full & 14.75 &   1.91  &   0.29  &   2.18  \\
DD2 1.2-1.2          & full & 13.21 &  3.09   &  0.17   &   1.37  \\
DD2 1.2-1.35         & full & 13.21 &  3.17   &  0.20   &   2.06  \\
DD2 1.2-1.6          & full & 13.21 & 10.90   &  0.20   &   6.39  \\
DD2 1.2-1.8          & full & 13.21 & 17.08   &  0.17   &   6.72  \\
DD2 1.35-1.5         & full & 13.21 &  3.57   &  0.25   &   3.13  \\
DD2 1.35-1.8         & full & 13.21 & 14.85   &  0.21   &   9.48  \\
DD2 1.5-1.5          & full & 13.21 &  5.38   &  0.26   &   4.66  \\
DD2 1.5-1.8          & full & 13.21 & 18.84   &  0.25   &  15.52  \\
DD2 1.6-1.6          & full & 13.21 &  7.80   &  0.27   &   7.40  \\
DD2 1.8-1.8          & full & 13.21 &  1.37   &  0.26   &   1.63  \\
DD2 1.35-2.0         & full & 13.21 &  6.41   &  0.31   &   9.64  \\
DD2 2.0-2.0          & full & 13.21 &  0.25   &  0.25   &   0.25  \\
SFHO 1.2-1.2         & full & 11.92 &   1.88  &  0.21   &   1.26  \\
SFHO 1.2-1.35        & full & 11.92 &   5.44  &  0.22   &   3.86  \\
SFHO 1.2-1.6         & full & 11.92 &  16.91  &  0.21   &  11.10  \\
SFHO 1.2-1.8         & full & 11.92 &   5.78  &  0.34   &  10.08  \\
SFHO 1.35-1.5        & full & 11.92 &  18.73  &  0.23   &  13.34  \\
SFHO 1.35-1.8        & full & 11.92 &  11.76  &  0.31   &  16.22  \\
SFHO 1.5-1.5         & full & 11.92 &   4.10  &  0.27   &   4.13  \\
SFHO 1.5-1.8         & full & 11.92 &   6.34  &  0.42   &  14.40  \\
SFHO 1.6-1.6         & full & 11.92 &   1.13  &  0.21   &   1.00  \\
SFHO 1.8-1.8         & full & 11.92 &   0.17  &  0.29   &   0.24  \\
\tableline
 \end{tabular}
\tablecomments{Basic properties of simulations and the employed EoSs. In the second column the thermal ideal-gas index is given if thermal effects are incorporated approximately; ``full'' refers to a simulation with a fully consistent treatment of thermal effects. $R_{1.35}$ specifies the radius of a nonrotating NS with 1.35~$M_{\odot}$. $M_{\mathrm{ej}}$ is the amount of unbound matter. The fifth column provides the average outflow velocity. $E_{\mathrm{kin}}$ denotes the kinetic energy of the ejecta. A cross in the first column indicates EoSs which are incompatible with the observation of a $(2.01\pm 0.04)~M_{\odot}$ NS~\citep{Antoniadis26042013}.}
           
 \end{ruledtabular} 
\end{table}
           
\begin{figure*}
\includegraphics[width=8.9cm]{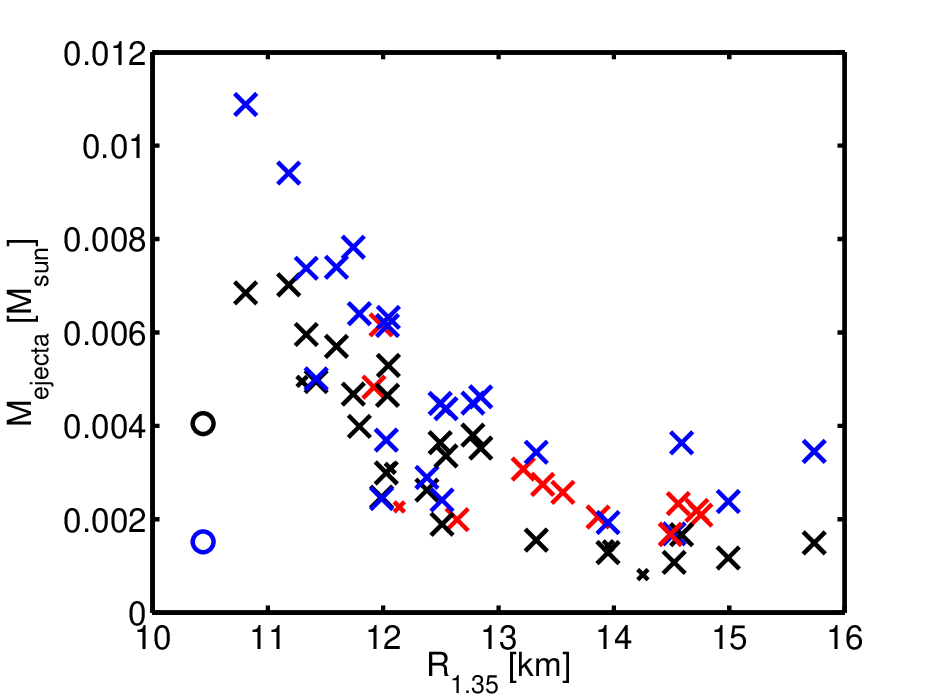}
\includegraphics[width=8.9cm]{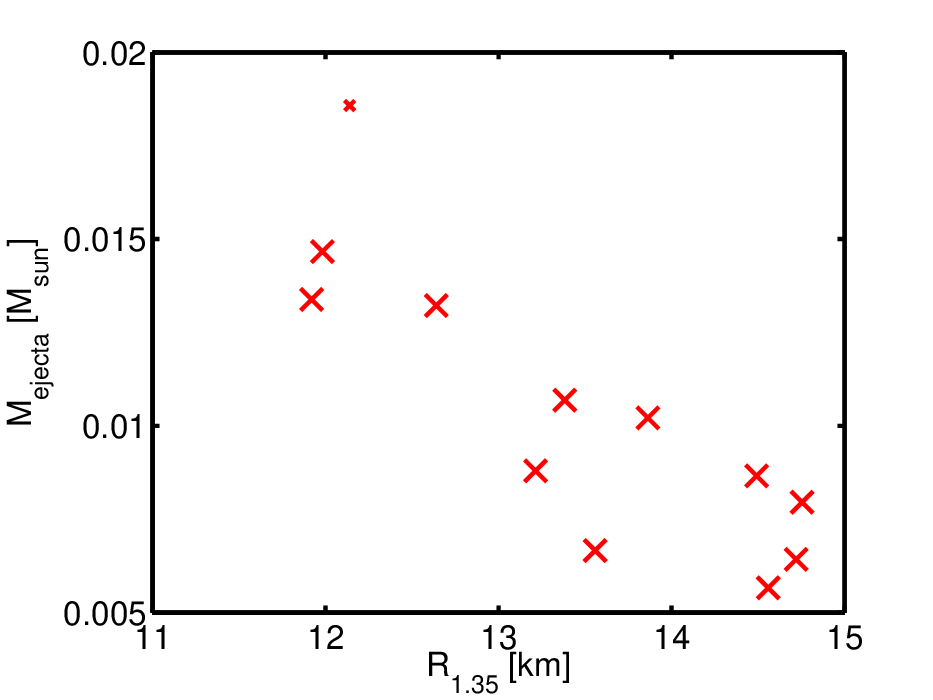}
           
\includegraphics[width=8.9cm]{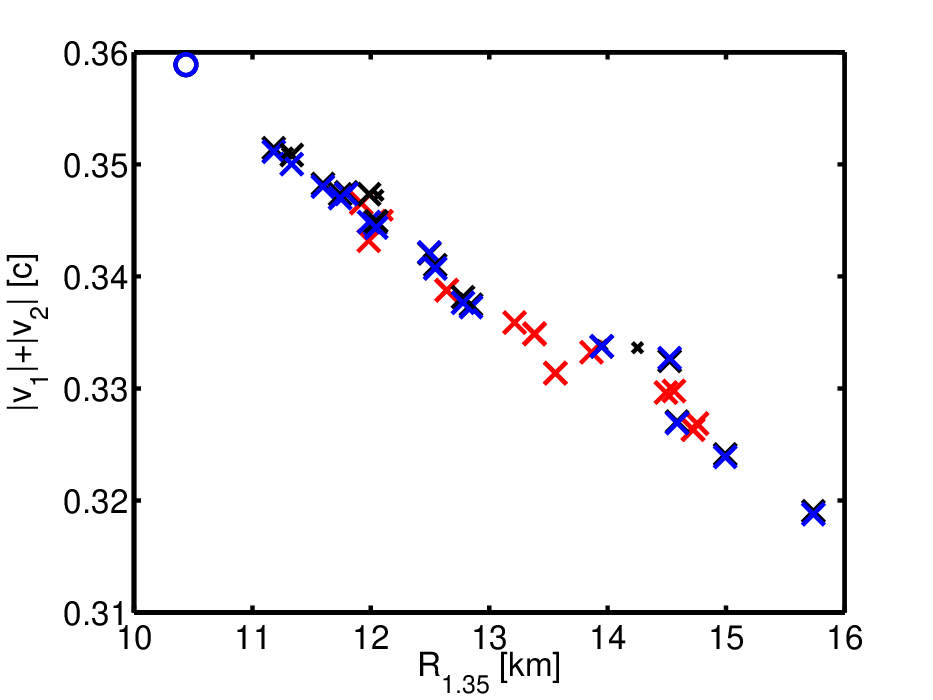}
\includegraphics[width=8.9cm]{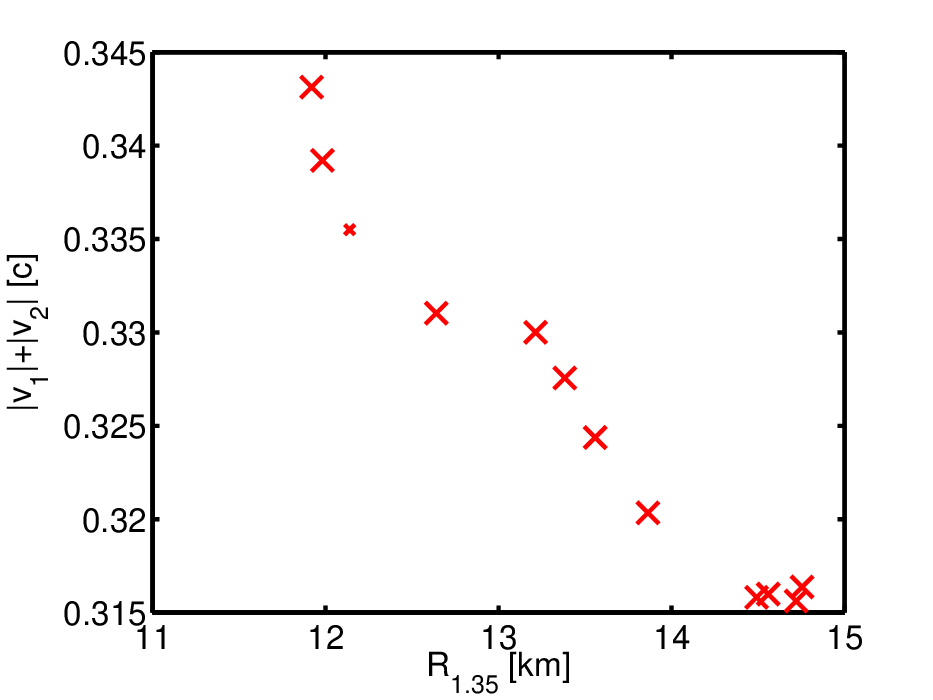}
\caption{\label{fig:mejr135}Amount of unbound material for 1.35-1.35~$M_{\odot}$ mergers (top left) and 1.2-1.5~$M_{\odot}$ mergers (top right) for different EoSs characterized by the corresponding radius $R_{1.35}$ of a nonrotating NS. Red crosses denote EoSs which include thermal effects consistently, while black (blue) symbols indicate zero-temperature EoSs that are supplemented by a thermal ideal-gas component with $\Gamma_{\mathrm{th}}=2$ ($\Gamma_{\mathrm{th}}=1.5$) (see main text). Small symbols represent EoSs which are incompatible with current NS mass measurements~\citep{2010Natur.467.1081D,Antoniadis26042013}. Circles display EoSs which lead to the prompt collapse to a black hole. The lower panels display the sum of the maxima of the coordinate velocities of the mass centers of the two binary components as a function of $R_{1.35}$ for symmetric (bottom left) and asymmetric (bottom right) binaries.}
\end{figure*}

The upper left panel of Fig.~\ref{fig:mejr135} displays the amount of unbound material as a function of $R_{1.35}$ for all 40 EoSs used in our study (see also Table~\ref{tab:models}). Red crosses identify EoSs which provide the full temperature dependence. The black symbols correspond to barotropic zero-temperature EoSs, which are supplemented by a thermal ideal-gas component choosing $\Gamma_{\mathrm{th}}=2$ (see Sect.~\ref{sec:code}). Results based on the same zero-temperature EoS but with $\Gamma_{\mathrm{th}}=1.5$ are given in blue at the same radius $R_{1.35}$. Small symbols indicate results for EoSs which are excluded by the pulsar mass measurement of~\citet{Antoniadis26042013}. Circles mark cases which lead to the prompt collapse to a black hole.

One can recognize a clear EoS dependence of the ejecta mass, where EoSs with a high compactness of the NSs lead to an enhanced production of unbound material. The ejecta mass can be as big as about 0.01~$M_\odot$ for symmetric mergers with a total binary mass $M_{\mathrm{tot}}=M_1+M_2=2.7~M_{\odot}$. EoSs with relatively large NS radii lead to outflow masses of about 0.001 to 0.002~$M_\odot$. For EoSs with approximately the same $R_{1.35}$ the ejecta masses show a scatter of up to 0.003~$M_{\odot}$. However, considering only EoSs with a fully consistent description of thermal effects (red symbols) the variations are smaller. Only one simplified EoS (eosAU) leads to a prompt collapse of the merger remnant and yields significantly smaller ejecta masses (circles). Using the radius $R_{1.6}$ of a nonrotating NS with 1.6~$M_{\odot}$ or the radius $R_{\mathrm{max}}$ of the maximum-mass Tolman-Oppenheimer-Volkoff solution to characterize an EoS results in diagrams similar to the upper left panel of Fig.~\ref{fig:mejr135}. However, no clear trend can be found for the ejecta mass as a function of the maximum mass of nonrotating NSs. We therefore conclude that the NS compactness is the crucial EoS parameter determining the ejecta mass. Indications of such a behavior were already observed in simulations for four simplified EoSs with an approximate temperature treatment~\citep{2012arXiv1212.0905H}.

The dynamics of the merger explain why small NS radii lead to higher ejecta masses. For smaller $R_{1.35}$ the inspiral phase lasts longer and the stars reach higher velocities before they collide. The relation between the impact velocity and the NS radius is clearly seen in the lower left panel of Fig.~\ref{fig:mejr135}, which displays the sum of the maxima of the coordinate velocities of the mass centers of the two binary components. The maximum of the coordinate velocity is reached shortly after the first contact, before the cores of the NSs are decelerated by the collision. The clash of more compact NSs is more violent and more material is squeezed out from the collision interface, for which reason the negative correlation of the velocities with $R_{1.35}$ is reflected by a similar negative correlation of $M_{\mathrm{ejecta}}$ and $R_{1.35}$. Moreover, for smaller $R_{1.35}$ the central remnant consisting of the double cores rotates faster and the bounce and rebounce are stronger, i.e. the surface of the remnant moves faster and pushes away matter more efficiently.

\subsection{Influence of the approximate treatment of thermal effects}
A number of simulations of our survey (black and blue symbols) as well as calculations by other groups ~\citep[e.g.][]{2011ApJ...736L..21R,2012arXiv1212.0905H} rely on an approximate description of thermal effects in the EoS, which requires the specification of an effective thermal ideal-gas index (see Sect.~\ref{sec:code}). As can be seen in the upper left panel of Fig.~\ref{fig:mejr135}, the choice of the value for this ideal-gas index has a considerable impact on the ejecta mass; the simulations with $\Gamma_{\mathrm{th}}=1.5$ (blue symbols) yield generally more unbound matter. The reason is the reduced thermal pressure support, which means that the two dense cores can approach each other more closely during the collision, which results in a more violent impact and shearing motion and thus in more material being squeezed out from the collision interface and in a more powerful oscillation of the central remnant. This can be clearly seen by following the centers of mass of the two cores or the evolution of the central lapse function.

The optimal choice of $\Gamma_{\mathrm{th}}$ is a priori unclear and may be different for different EoSs. To address this issue we performed additional simulations (not shown in Fig.~\ref{fig:mejr135}, but listed in Table~\ref{tab:models}) for temperature-dependent EoSs (SFHO, DD2, TM1, NL3) after reducing them to the zero-temperature sector (with the constraint of neutrino-less beta-equilibrium) and supplementing them with the approximate description of thermal effects using $\Gamma_{\mathrm{th}}=1.5,~1.8$ and 2. The comparison with the fully consistent simulations reveals that generally a choice of $\Gamma_{\mathrm{th}}=1.5$ yields the best quantitative agreement with only a slight underestimation of about 10 per cent (20 per cent for the SFHO EoS), whereas the ejecta masses with $\Gamma_{\mathrm{th}}=1.8$ or $\Gamma_{\mathrm{th}}=2$ are significantly too low compared to the fully consistent models (for the tested EoSs between 15 to 40 per cent for $\Gamma_{\mathrm{th}}=2$ and between 20 to 35 per cent for $\Gamma_{\mathrm{th}}=1.8$). 

The fact that a relatively low $\Gamma_{\mathrm{th}}$ reproduces the ejecta properties best contrasts the finding that a higher $\Gamma_{\mathrm{th}}$ (in the range between 1.5 and 2) has turned out to be more suitable for describing gravitational-wave features and the post-merger collapse behavior~\citep[see][]{2010PhRvD..82h4043B}, i.e. the bulk mass motion of the colliding stars. The reason for this discrepancy is the density dependence of $\Gamma_{\mathrm{th}}$, which drops from about 2 at supranuclear densities to about $4/3$ for densities below $\approx 10^{11}~\mathrm{g/cm^3}$ (see Fig.~2 in~\citet{2010PhRvD..82h4043B}). While the dynamics of the bulk mass of the merging objects, which is responsible for the gravitational-wave production, is fairly well captured with a choice of $\Gamma_{\mathrm{th}}\sim 2$,  unbound fluid elements originating from the inner crust, where most of the ejecta stem from, encounter different density regimes. Consequently, the ejecta behavior cannot be well modelled with the $\Gamma_{\mathrm{th}}$ that is appropriate for high-density matter. We found the best compromise to be $\Gamma_{\mathrm{th}}\approx 1.5$, but we stress that the results of simulations using an approximate treatment of thermal effects should be taken with caution and do not need to be quantitatively reliable in all aspects.
\subsection{Asymmetric binaries}\label{ssec:asym}
Even though most NS binaries are expected to be nearly symmetric systems with a total mass of about 2.7~$M_{\odot}$, we investigate the mass ejection of an asymmetric setup to check whether also in this case an EoS dependence exists. The upper right panel of Fig.~\ref{fig:mejr135} displays the ejecta masses for simulations of asymmetric binaries with a 1.2~$M_{\odot}$ NS and a 1.5~$M_{\odot}$ NS (see also Table~\ref{tab:models}). Again, the radius $R_{1.35}$ of a nonrotating NS is used to characterize different EoSs. Here we restrict ourselves to EoSs which provide the full temperature dependence. In comparison to the symmetric binary mergers the amount of unbound material is significantly larger. The ejecta masses are about a factor of two higher than for the symmetric binaries with the same total binary mass. Also for asymmetric binaries a decrease of $M_{\mathrm{ej}}$ with bigger $R_{1.35}$ is visible, but the scatter between models with similar $R_{1.35}$ is larger. The lower right panel of Fig.~\ref{fig:mejr135} shows the sum of the maxima of the coordinate velocities of the mass centers of the two asymmetric binary components. As in the symmetric case the two stars collide with a higher impact velocity if the initial radii of the NSs are smaller.

\begin{figure*}
\includegraphics[width=8.9cm]{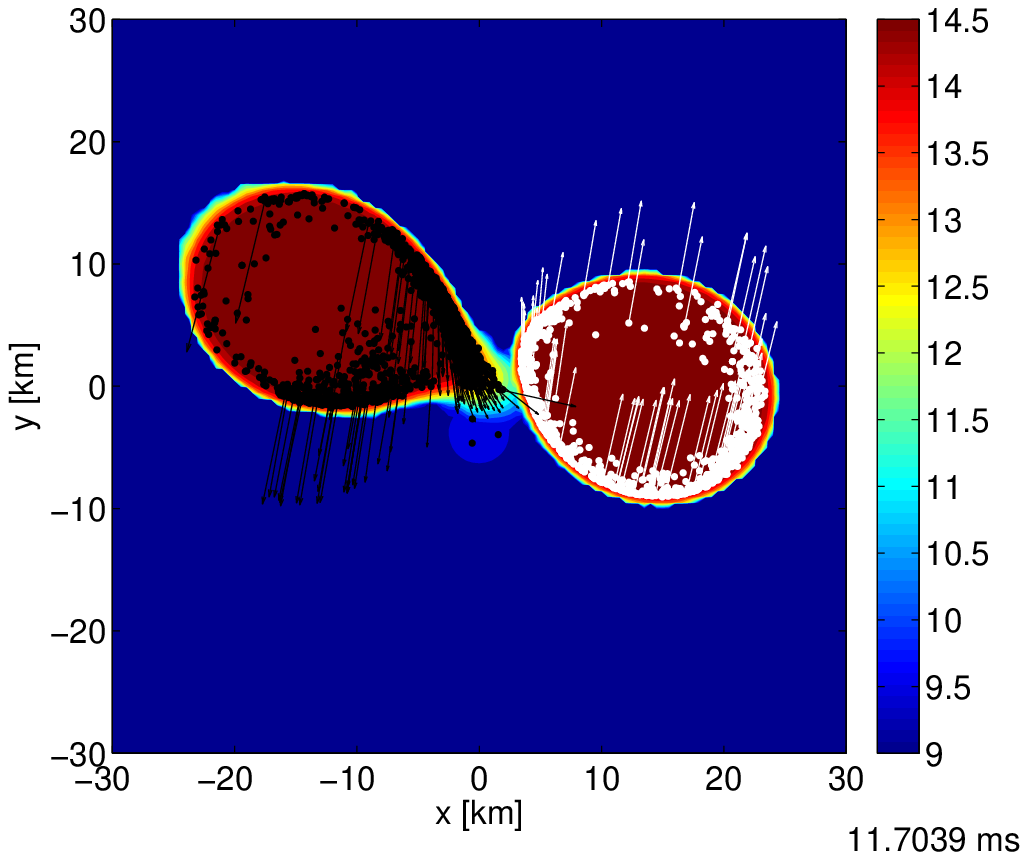}
\includegraphics[width=8.9cm]{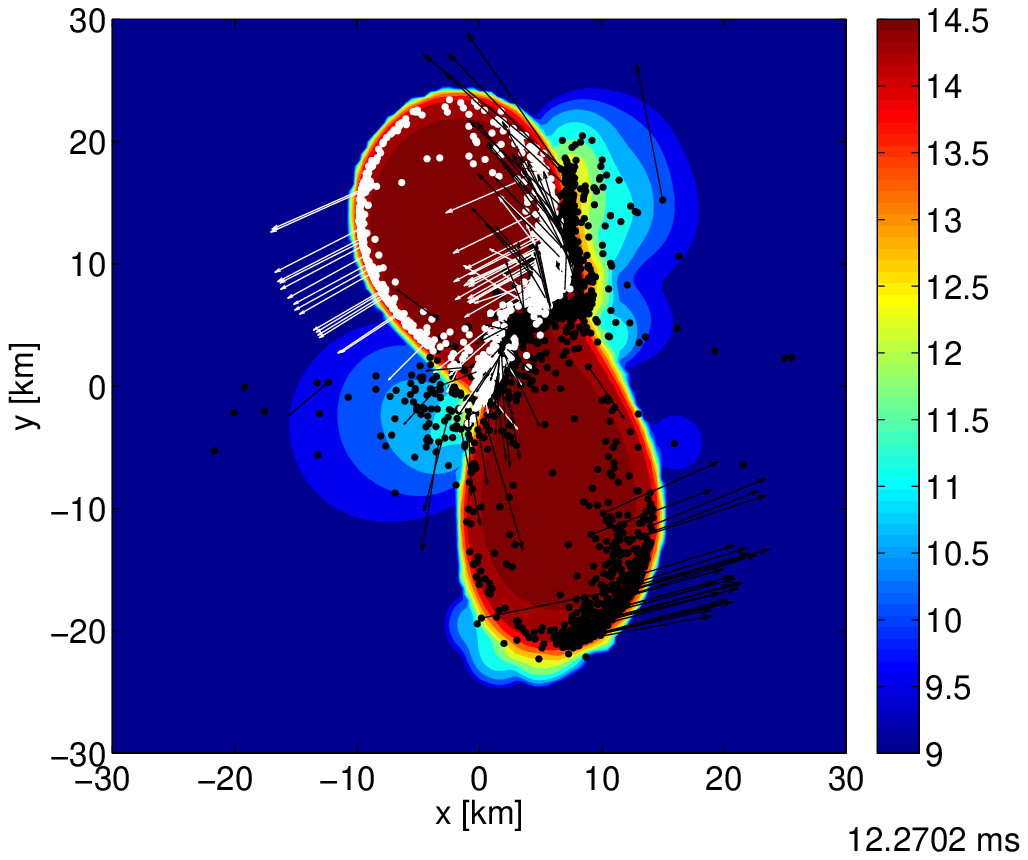}

\includegraphics[width=8.9cm]{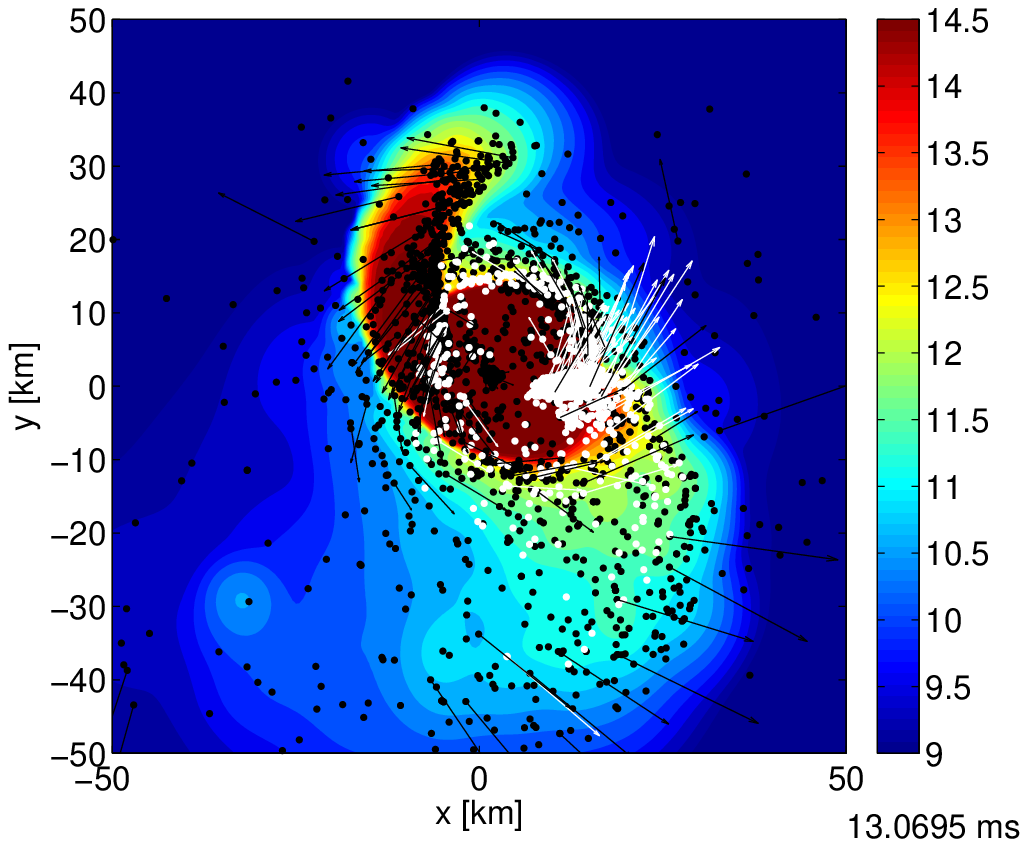}
\includegraphics[width=8.9cm]{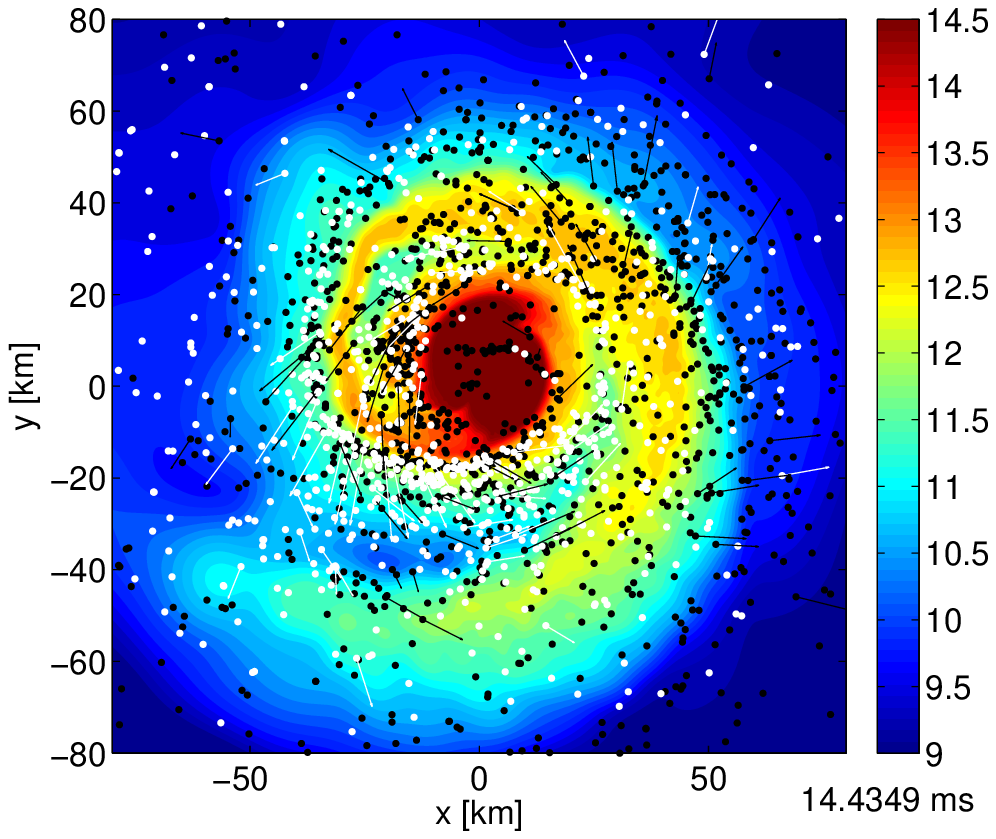}

\includegraphics[width=8.9cm]{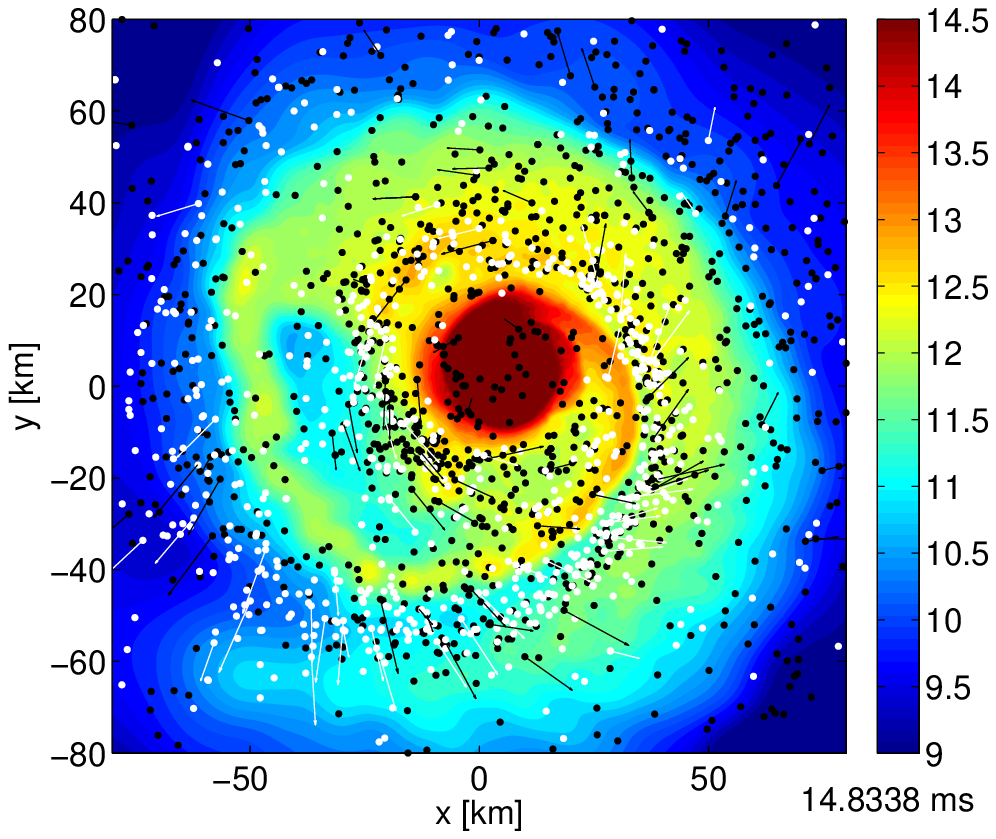}
\includegraphics[width=8.9cm]{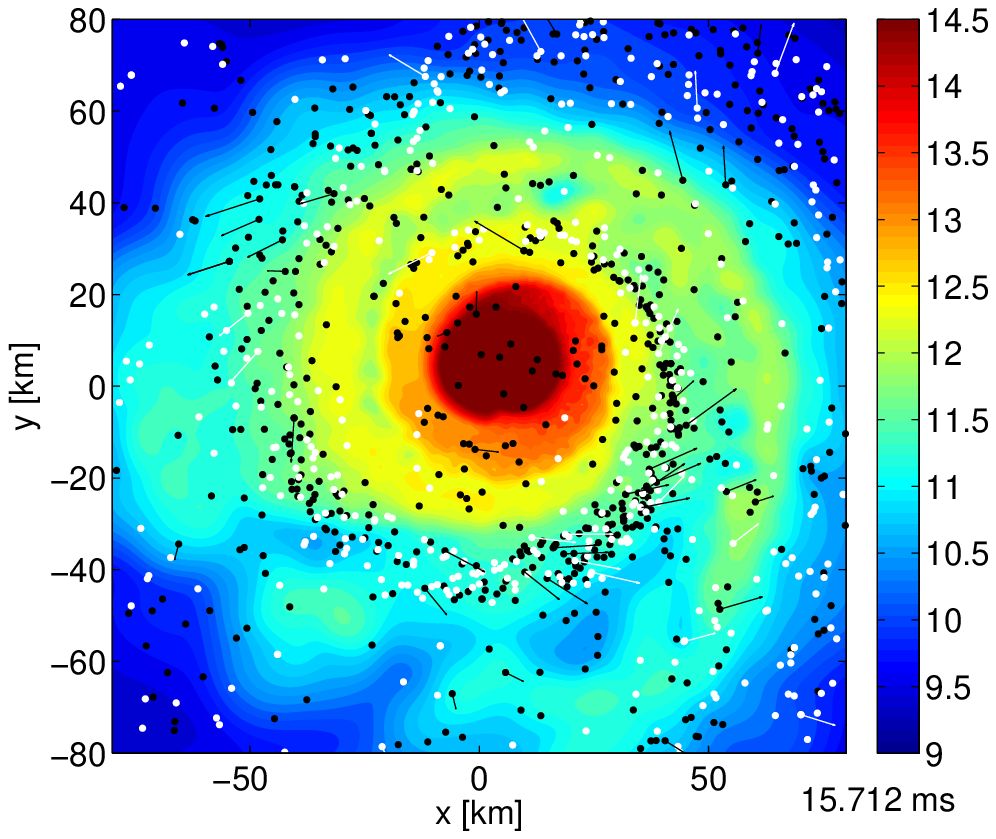}
\caption{\label{fig:snapasym}Same as Fig.~\ref{fig:snap}, but for the asymmetric 1.2-1.5~$M_{\odot}$ binary. Here the propagation speed of every 20th particle is indicated by an arrow and the side lengths of the panels differ from those of Fig.~\ref{fig:snap}. In the upper panels the lower-mass star is identified by the black particles. The visualization tool SPLASH was used to convert SPH data to grid data~\citep{2007PASA...24..159P}.}
\end{figure*}
\begin{figure*}
\includegraphics[width=8.9cm]{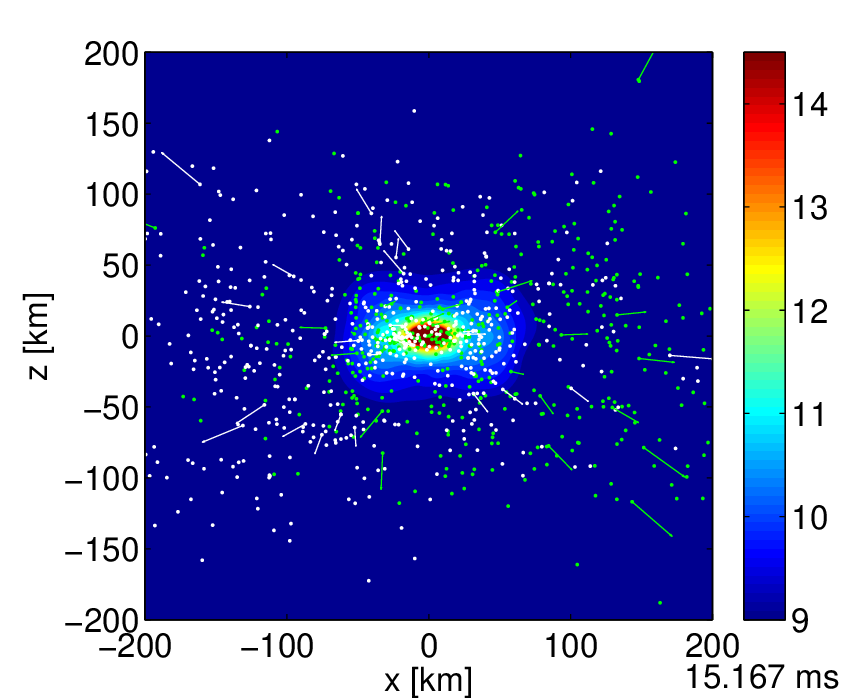}
\includegraphics[width=8.9cm]{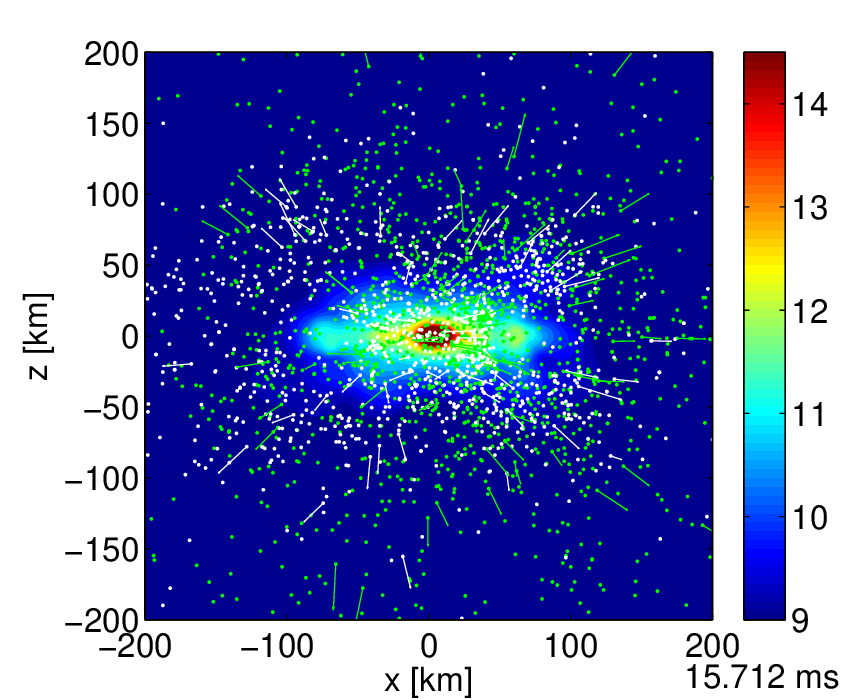}
\caption{\label{fig:snap_xz}Distribution of the conserved rest-mass density (color-coded and logarithmically plotted in $\mathrm{g/cm^3}$) in a plane perpendicular to the orbital plane for the 1.35-1.35~$M_{\odot}$ merger (left) and the 1.2-1.5~$M_{\odot}$ merger (right) with the DD2 EoS. The three-dimensional positions of unbound particles are projected into the cross-sectional plane, green and white referring to the origin from the one and the other NS. The velocity of every 20th particle is indicated by an arrwo. The time of the snapshots is given below the color bar of each panel. The visualization tool SPLASH was used to convert SPH data to grid data~\citep{2007PASA...24..159P}.}
\end{figure*}

Due to the asymmetry the dynamics of the merger proceeds differently from the symmetric case (see Fig.~\ref{fig:snapasym}). Prior to the merging the less massive binary component is deformed to a drop-like structure with the cusp pointing to the 1.5~$M_{\odot}$ NS (top panels). After the stars begin to touch each other, the lighter companion is stretched and a massive tidal tail forms (middle left panel). The deformed 1.2~$M_{\odot}$ component is wound around the more massive companion (middle panels). Also in the case of asymmetric mergers the majority of the ejecta originates from the contact interface of the collision, i.e. from the cusp of the ``tear drop'' and from the equatorial surface of the more massive companion, where the impact ablates matter (see top panels). Some matter at the tip of the cusp directly fulfills the ejecta criterion (top right panel), while the majority obtains an additional push by the interaction with the asymmetric, mass-shedding central remnant and the developing spiral arms (middle right and bottom panels). A smaller amount of ejecta of roughly 25 per cent originates from the outer end of the primary tidal tail (particles in the lower part of the top right panel). A part of this matter becomes unbound by tidal forces (at the tip of the tidal tail in the middle left panel) and  the other fraction by an interaction with the central remnant (middle left panel).

\begin{figure*}
\includegraphics[width=8.9cm]{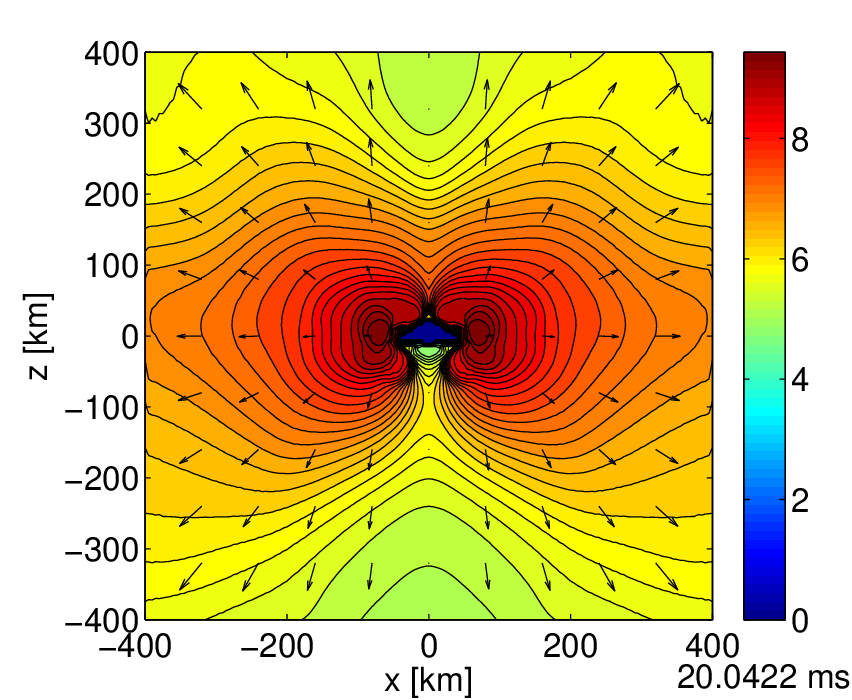}
\includegraphics[width=8.9cm]{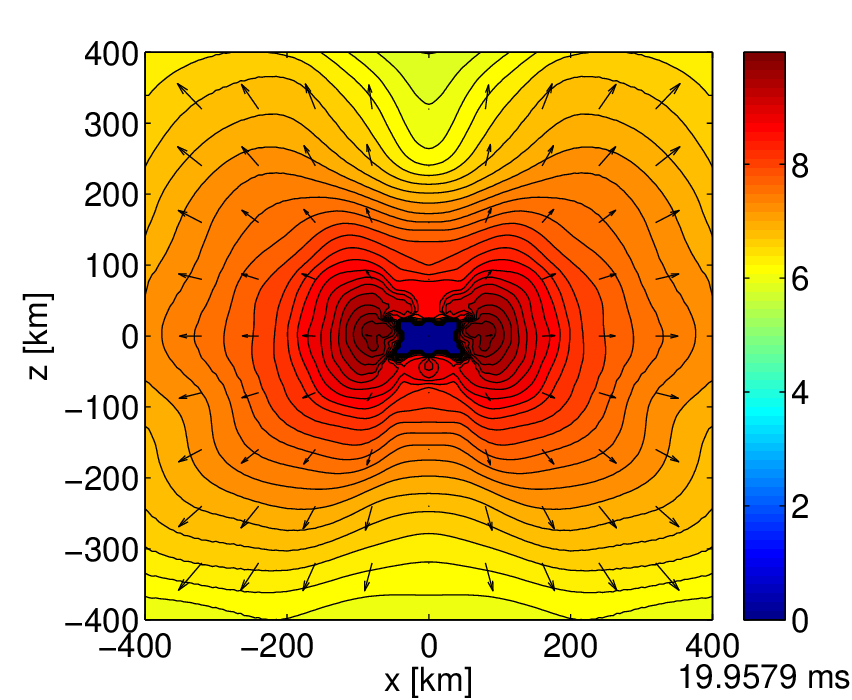}
\caption{\label{fig:ejectageo}Ejecta geometry visualized by the rest-mass density (color-coded and logarithmically plotted in $\mathrm{g/cm^3}$) excluding matter of the bound central remnant for the 1.35-1.35~$M_{\odot}$ merger (left) and the 1.2-1.5~$M_{\odot}$ merger (right) with the DD2 EoS. Density contours are obtained by azimuthal averaging. Arrows represent the coordinate velocity field where an arrow length of 200~km corresponds to the speed of light. The time of the snapshots is given below the color bar of each panel. The visualization tool SPLASH was used to convert SPH data to grid data~\citep{2007PASA...24..159P}.}
\end{figure*}

Figure~\ref{fig:snap_xz} displays the distribution of the ejecta in a plane perpendicular to the binary orbit for the symmetric merger (left panel) compared to the asymmetric merger (right panel) for the last timesteps shown in Fig.~\ref{fig:snap} and Fig.~\ref{fig:snapasym}, respectively. A considerable fraction of the ejected matter is expelled with large direction angles relative to the orbital plane. For a timestep about 5~ms later the ejecta geometry is visualized (azimuthally averaged) in Fig.~\ref{fig:ejectageo} excluding the bound matter. For both mergers the outflows exhibit a (torus or donut-like) anisotropy with an axis ratio of about 2:3. The velocity fields also show a slight dependence on the direction.

\subsection{Binary parameter dependence}\label{ssec:binpara}
The exploration of the full space of possible binary parameters is interesting for the determination of the highest and lowest possible ejecta mass for a given EoS and to understand the influence of the binary setup on the ejecta production. Such an investigation has been conducted only for one EoS (Shen) by Newtonian calculations~\citep{2012arXiv1206.2379K,2012arXiv1204.6240R,2012arXiv1210.6549R} and within a relativistic framework~\citep{2007A&A...467..395O}, which revealed quantitative differences between both approaches. Other surveys using different EoSs have been restricted to a limited variation of the binary masses~\citep{2011ApJ...736L..21R,2011ApJ...738L..32G,2012arXiv1212.0905H}. Here we present the dependence of the ejecta mass on the mass ratio $q=M_1/M_2$ and the total binary mass $M_{\mathrm{tot}}=M_1+M_2$ for a subset of EoSs employed in our study. The NL3, DD2 and SFHO EoSs are chosen because they are representative for the full set of possible EoSs: While the NL3 EoS is relatively stiff, resulting in $R_{1.35}=14.75$~km, the soft SFHO EoS produces rather compact NSs with $R_{1.35}=11.74$~km, and the DD2 represents an intermediate case with $R_{1.35}=13.21$~km.
\begin{figure}
\includegraphics[width=8.9cm]{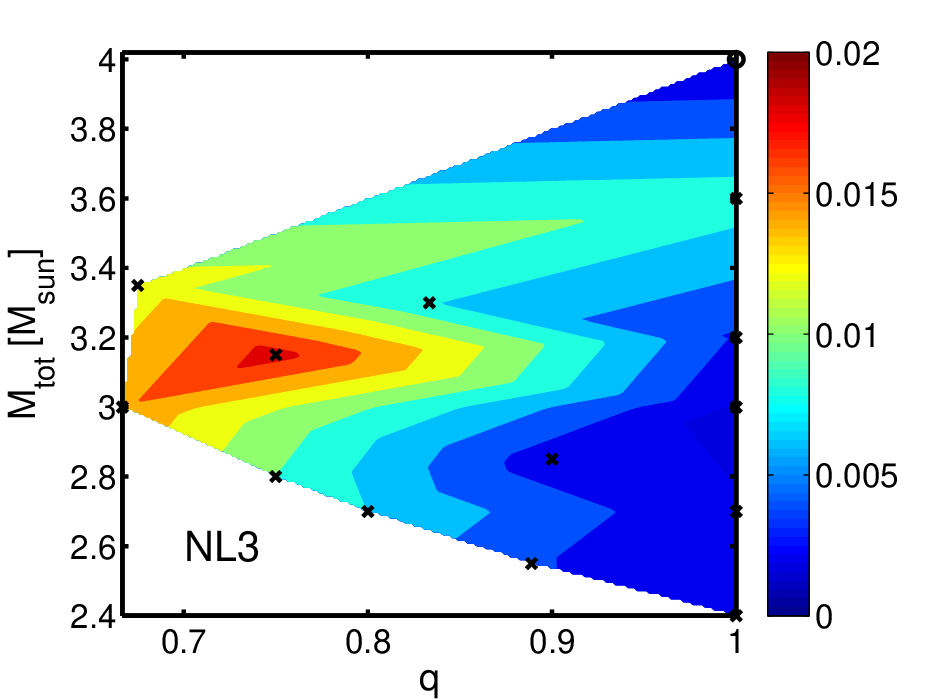}
\includegraphics[width=8.9cm]{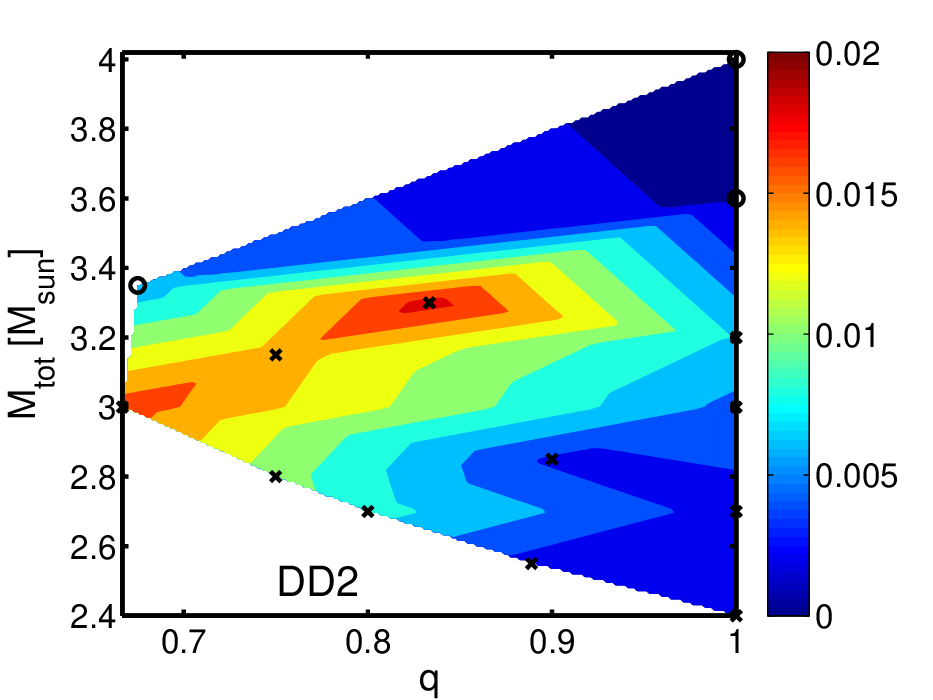}
\includegraphics[width=8.9cm]{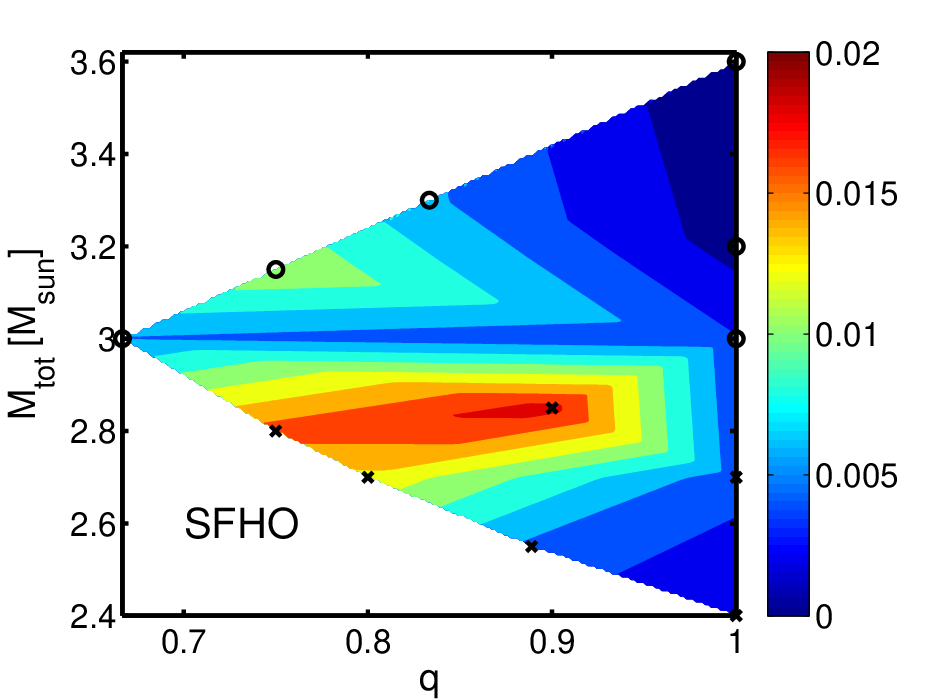}
\caption{\label{fig:binpara3eos}Ejecta mass in $M_{\odot}$ as function of the mass ratio $q$ and the total binary mass $M_{\mathrm{tot}}$ for the soft SFHO EoS (upper panel), the intermediate DD2 EoS (center panel) and the stiff NL3 EoS (bottom panel). In all panels the simulated binary setups are marked by symbols. Crosses indicate the formation of a differentially rotating NS remnant, while circles identify configurations which lead to a direct gravitational collapse.}
\end{figure}
Figure~\ref{fig:binpara3eos} displays the amount of unbound matter as a function of the mass ratio $q$ and total binary mass $M_{\mathrm{tot}}$ for these three EoSs (see also Table~\ref{tab:models}). The simulated binary configurations that form a differentially rotating NS are indicated in the figures by crosses, whereas systems leading to prompt black-hole formation (within about one millisecond after the first contact) are marked with circles (the ejecta properties are extracted 10~ms after the merging).

All three EoSs show qualitatively the same behavior. A clear trend of increasing ejecta masses with larger binary asymmetry is visible. For symmetric binaries that do not undergo gravitational collapse there is also a slight tendency of higher total binary masses leading to more unbound material. The increase with $M_{\mathrm{tot}}$ is more pronounced for asymmetric systems. The occurrence of a prompt collapse results in a significant drop of the ejecta mass. This is an important qualitative difference to Newtonian calculations, which cannot determine and follow the relativistic gravitational collapse. The threshold for the prompt collapse depends sensitively on the EoS, and soft EoSs lead to a collapse for relatively small $M_{\mathrm{tot}}$.

For all EoSs the maximum ejecta masses (in all cases slightly below 0.02~$M_{\odot}$) are four to ten times higher than the amount of unbound matter of the symmetric 1.35-1.35~$M_{\odot}$ binaries. Here, the absolute differences between the maximum and the minimum ejected mass for non-collapsing cases are larger for stiff EoSs like the NL3 and less pronounced for soft EoSs like the SFHO. Soft EoSs yield steeper gradients of the ejecta mass in the binary parameter space, i.e. a certain variation in the binary parameters leads to a larger change of the ejecta masses than it is the case for a stiff EoS. To a good approximation and ignoring cases with a prompt black hole formation, the setup with two stars of about 1.35~$M_{\odot}$ is the system that produces the smallest amount of ejecta for the majority of investigated EoSs. The SFHO EoS is one of the exceptions, for which, e.g., the 1.2-1.2~$M_{\odot}$ binary yields a factor two to three less ejecta than the 1.35-1.35~$M_{\odot}$ setup.

Based on Newtonian calculations a fit formula for the ejecta mass as fraction of $M_{\mathrm{tot}}$ was proposed as a function of $\eta=1-4M_1M_2/(M_1+M_2)^2$ in~\citet{2012arXiv1206.2379K} and~\citet{2012arXiv1210.6549R}. Reviewing our data (even without the prompt collapse cases) we find a more complicated behavior and we can neither confirm the validity of the suggested fit formula nor find a generalization of it. This is not unexpected in view of the quantitative and qualitative differences between Newtonian and relativistic simulations discussed above.

\subsection{Folding with binary populations}\label{ssec:fold}
The dependence of the ejecta mass on the binary parameters is essential to determine the total amount of ejecta produced by the binary population within a certain time and thus to estimate the average amount of ejecta per merger event. The properties of the NS binary population are provided by theoretical binary evolution models, which still contain considerable uncertainties in many complexities of single star evolution and binary interaction. Using the standard model of~\citet{2012ApJ...759...52D} 
the folding of our results with the binary population yields an average ejecta mass per merger event of about $3.6\times 10^{-3}~M_{\odot}$ for the NL3 EoS, $3.2\times 10^{-3}~M_{\odot}$ for the DD2 EoS, and $4.3\times 10^{-3}~M_{\odot}$ for the SFHO EoS. Therefore, the ejecta masses of the 1.35-1.35~$M_{\odot}$ binary mergers give numbers for the three cases which approximate the average amount of ejecta per merger event quite well (within 70 per cent for NL3, 3 per cent for DD2, 11 per cent for SFHO). This finding is simply a consequence of the fact that the binary distribution is strongly peaked around nearly symmetric systems with $M_{\mathrm{tot}}\approx 2.5~M_{\odot}$ so that the average ejecta mass is not sensitive to the larger ejecta production of asymmetric systems in the suppressed wings of the binary distribution.

\section{Nucleosynthesis}\label{sec:nucleo}
\subsection{R-process abundances}
The potential of NS mergers to produce heavy r-process elements in their ejecta has been manifested by several studies based on hydrodynamical simulations~\citep{1999ApJ...525L.121F,2010MNRAS.406.2650M,2011ApJ...736L..21R,2011ApJ...738L..32G,2012arXiv1206.2379K}. These investigations have considered only a few high-density EoSs (two EoSs were used in~\citet{2011ApJ...738L..32G}). Since the NS EoS affects sensitively the dynamics of NS mergers and thus the properties of the ejecta (amount, expansion velocity, electron fraction, temperature), we explore here the influence of the NS EoS on the r-process nucleosynthesis in a systematic way.

For a selected, representative set of EoSs we extract the thermodynamical histories of fluid elements which get gravitationally unbound. For these trajectories nuclear network calculations were performed as in~\citet{2011ApJ...738L..32G}, where details on the reaction network, the temperature postprocessing and the density extrapolation beyond the end of the hydrodynamical simulations can be found. The reaction network includes all 5000 species from protons up to Z=110 lying  between the valley of $\beta$-stability and the neutron-drip line. All fusion 
reactions on light elements, as well as radiative neutron captures, photodisintegrations, $\beta$-decays and fission processes are included. The corresponding rates are based on experimental data whenever available or on theoretical predictions otherwise, as prescribed in the BRUSLIB nuclear astrophysics library~\citep{2013AA.549.A110}

\begin{figure}
\includegraphics[width=8.9cm]{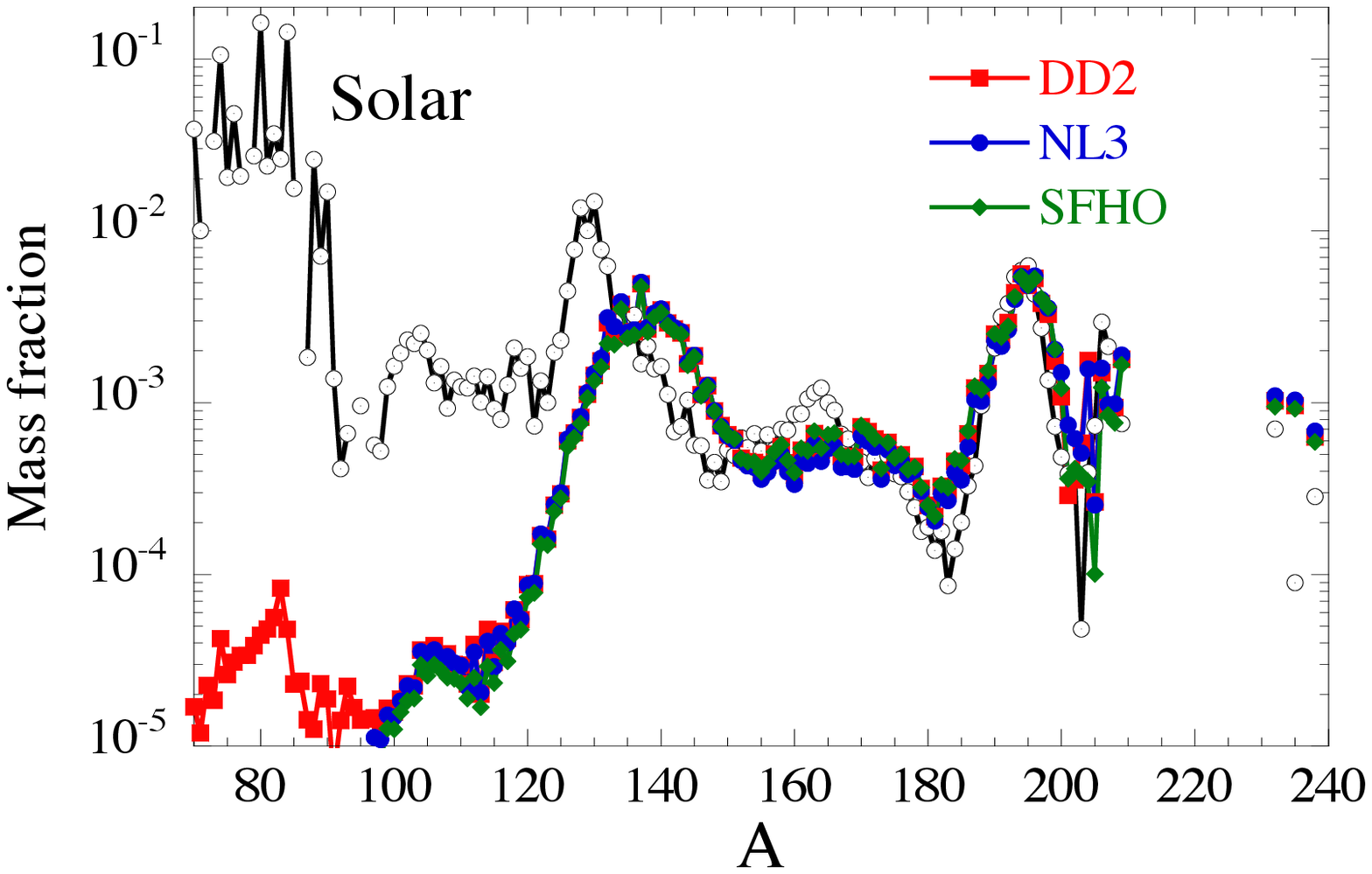}
\caption{\label{fig:abund135135}Nuclear abundance pattern for the 1.35-1.35~$M_{\odot}$ mergers with the NL3 (blue), DD2 (red) and SFHO (green) EoSs compared to the solar r-process abundance distribution (black).}
\end{figure}

Figure~\ref{fig:abund135135} shows the final nuclear abundance patterns for the 1.35-1.35~$M_{\odot}$ mergers described by the NL3 (blue), DD2 (red) and SFHO (green) EoSs. For every model about 200 trajectories were processed, which roughly correspond to about one tenth of the total ejecta. Comparing the final abundance distributions of the DD2 EoS for about 200 and the full set of 1000 fluid-element histories reveals a very good quantitative agreement, which proves that a properly chosen sample of about 200 trajectories is sufficient to be representative for the total amount of unbound matter.
\begin{figure}
\includegraphics[width=8.9cm]{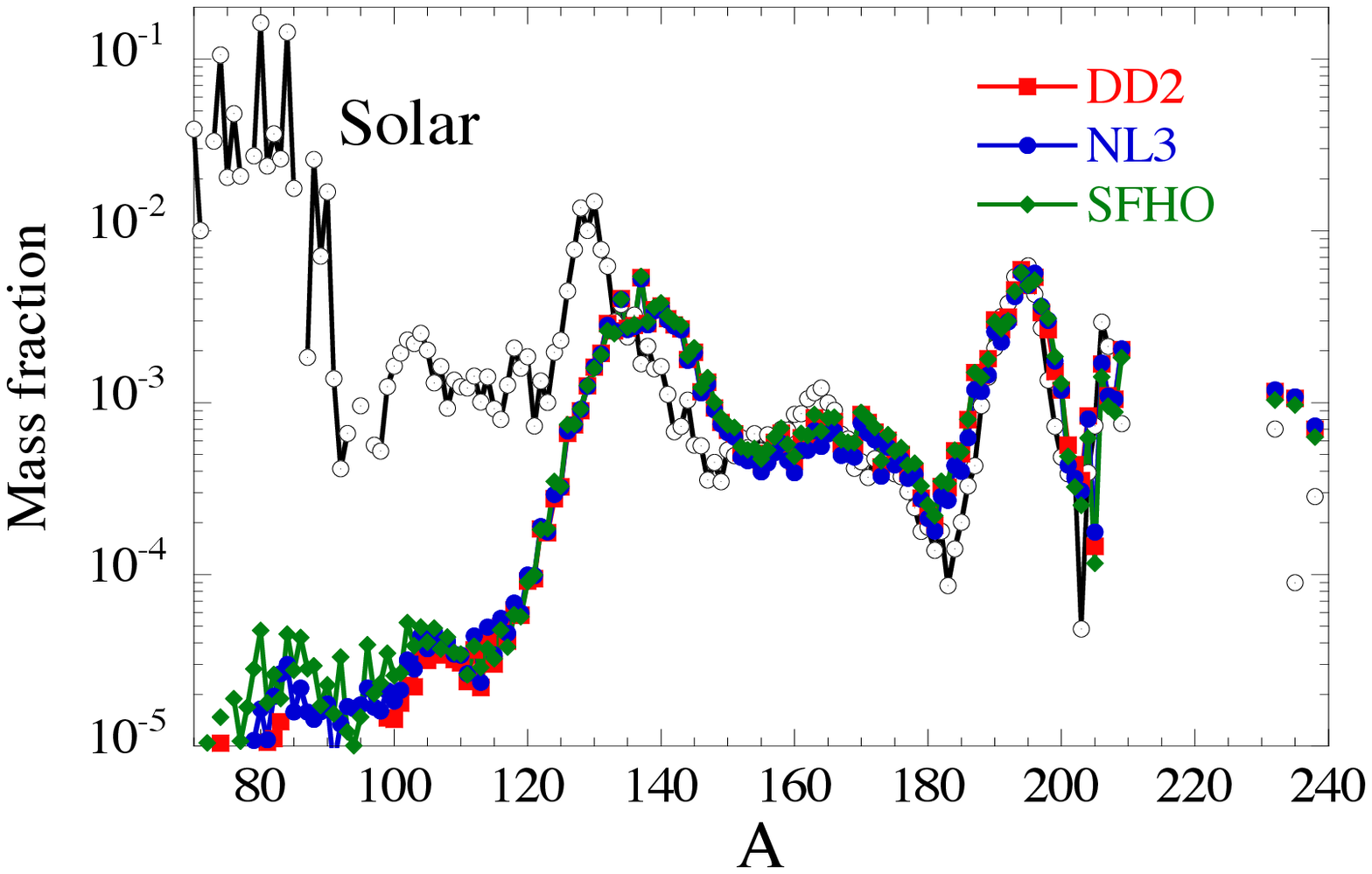}
\caption{\label{fig:abund1215}Nuclear abundance pattern for the 1.2-1.5~$M_{\odot}$ mergers with the NL3 (blue), DD2 (red) and SFHO (green) EoSs compared to the solar r-process abundance distribution (black).}
\end{figure}

The scaled abundance patterns displayed in Fig.~\ref{fig:abund135135} match closely the solar r-process composition above mass number $A\approx 140$. In particular the third r-process peak around $A=195$ is robustly reproduced by all models. Above mass number $A\approx 100$ the results for the different NS EoSs hardly differ. For all three displayed models the peak around $A\approx 140$ is produced by fission recycling, which occurs when the nuclear flow reaches fissioning nuclei around $^{280}$No at the end of the neutron irradiation during the $\beta$-decay cascade.
The exact shape and location of this peak are therefore strongly affected by the theoretical modeling of the fission processes (including in particular the fission fragment distribution of the fissioning nuclei) which are still subject to large uncertainties~\citep{go09}. Hence, the deviations from the solar abundance pattern between $A\approx 130$ and $A \approx 170$ are not unexpected, while the third r-process peak around $A=195$ is a consequence of the closed neutron shell at $N=126$, which is robustly predicted by theoretical models. Very similar results were obtained for NS merger models performed with the LS220 and Shen EoSs in~\citet{2011ApJ...738L..32G}.

In Fig.~\ref{fig:abund1215} the normalized abundance patterns are shown for asymmetric 1.2-1.5~$M_{\odot}$ mergers employing the same representative EoSs as in Fig.~\ref{fig:abund135135}. Again a very good agreement between the solar r-process abundances and the calculated element distributions above $A\approx 130$ is found for all three high-density EoSs. This confirms earlier findings that the binary mass ratio has a negligible effect on the abundance yield distribution~\citep{2011ApJ...738L..32G}. It also confirms that the ejected abundance distribution is rather insensitive to the adopted EoS.

Besides the three temperature-dependent EoSs considered above we conducted network calculations also for merger models computed with zero-temperature EoSs supplemented by an approximate treatment of thermal effects with a $\Gamma_{\mathrm{th}}=2$ ideal-gas component (BSk20, BSk21). In this case the temperature is estimated following a procedure described in~\citet{2008PhRvD..77h4002E}, which converts the specific internal energy to temperature values. Doing so it is assumed that the energy of the thermal ideal-gas component is composed of the thermal energy of an ideal nucleon gas and a contribution from ultrarelativistic particles (photon, possibly electrons, positrons and neutrinos).
\begin{figure}
\includegraphics[width=8.9cm]{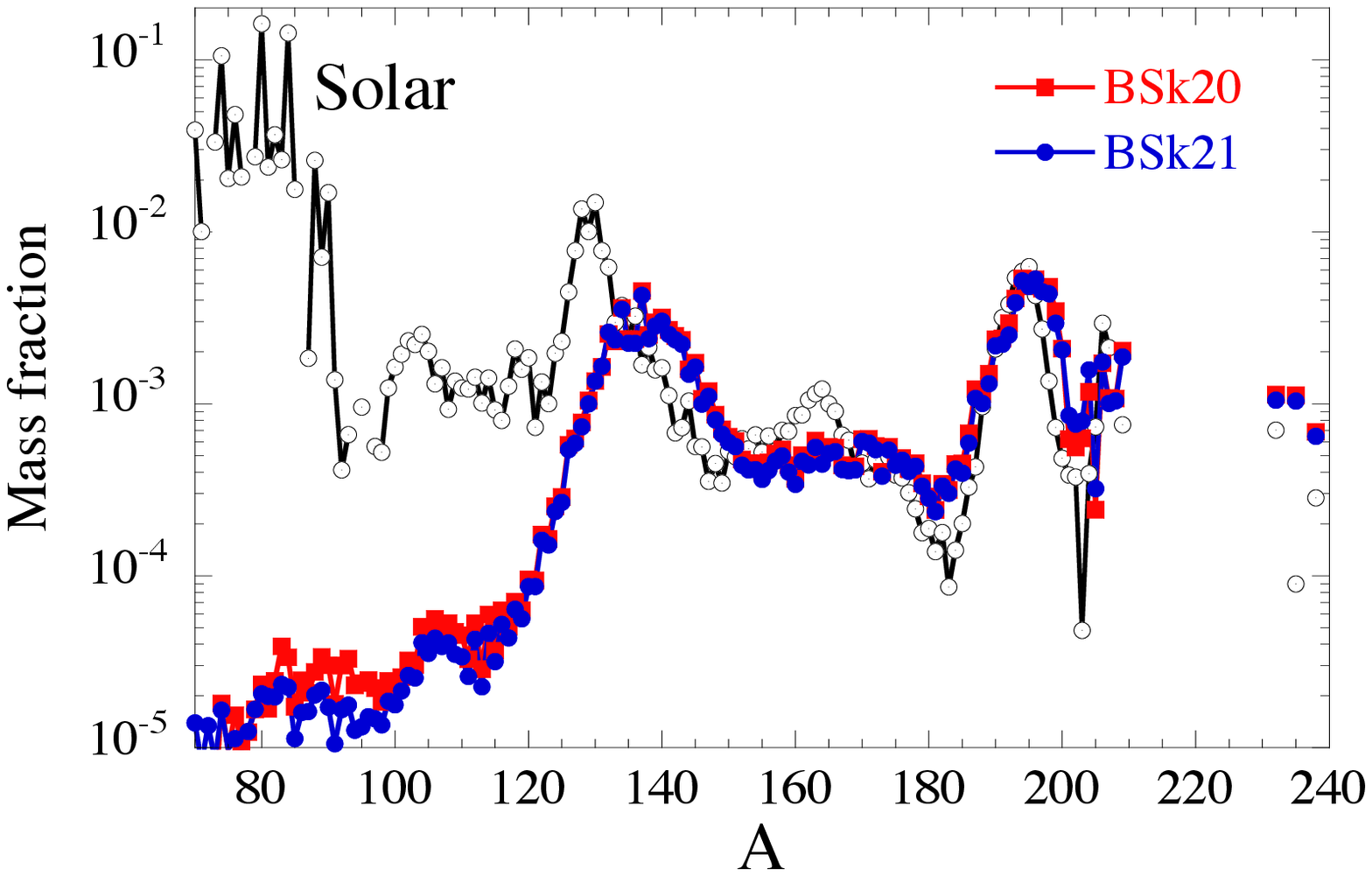}
\caption{\label{fig:abundbsk}Nuclear abundance pattern for the 1.35-1.35~$M_{\odot}$ mergers with the BSk20 (red) and BSk21 (blue) EoSs compared to the solar r-process abundance distribution (black).}
\end{figure}

The network calculations for the BSk20 and BSk21 EoSs yield an abundance pattern above $A\approx 130$ very similar to the other fully temperature-dependent EoSs (see Fig.~\ref{fig:abundbsk}). Differences between the fully consistent models and the simulations with approximate temperature treatment are found below mass number $A\approx 50$, where the calculations with the BSk EoSs yield a lower amount of elements with $5<A<50$ but a higher mass fraction of hydrogen, deuterium and helium. The reason is the higher temperatures found with the BSk EoSs at the beginning of the network calculations which lead to a reduced recombination of nucleons and $\alpha$-particles and consequently a smaller production of heavier nuclei. In this respect the conditions in the outflows of these models resemble the situation in the neutrino-driven winds of core-collapse supernovae but for significantly higher neutron excesses.

Overall, it is reassuring that the BSk models yield a similar abundance pattern of r-process elements, although these calculations rely on an approximate incorporation of thermal effects and a rough estimate of temperatures. It is an important finding of our work that r-process elements are robustly produced for a representative, diverse sample of high-density EoSs and that the outcome is insensitive to the exact initial temperature conditions and the binary setup.

\subsection{Merger rates}
The above-mentioned variations in the production of light elements are also reflected in the fraction of the ejecta which end up as r-process elements. In Tab.~\ref{tab:nucleo} a clear difference is observed for temperature-dependent EoSs (NL3, DD2, SFHO) and the BSk EoSs with approximate temperature treatment. While in the former cases about 96 to 99 per cent of the ejecta are converted to r-process elements, only 93 to 95 per cent of the ejecta are processed to r-process material in the latter models. This variation is a consequence of the temperature history affecting the production of light nuclei, as discussed above.

Despite these differences,  it is justified to assume that almost the total amount of ejecta is converted to heavy r-process elements, also considering the uncertainties in the determination of the exact ejected masses from simulations. Furthermore, in Sect.~\ref{sec:masses} it was argued that the total amount of ejecta, and thus the r-process material produced per event by the population of NS binaries, can be well represented by the yield of the 1.35-1.35~$M_{\odot}$ merger. This allows an important consistency check by comparing the theoretically expected production to the observed amount of r-process matter with $A\gtrsim 130$ in the Galaxy, which is estimated to be about $4\times 10^3~M_{\odot}$~\citep{2000ApJ...534L..67Q}. In order to produce this amount of heavy r-process elements within the Galactic history for about $10^{10}$~yr, one requires a merger rate of $4\times 10^{-4}/\mathrm{yr}$ if every coalescence ejects on average $10^{-3}~M_{\odot}$, which in our EoS survey corresponds to the lower bound on the ejecta mass of 1.35-1.35~$M_{\odot}$ mergers (see upper left panel of Fig.~\ref{fig:mejr135}). Similarly, assuming that NS mergers are the dominant source of heavy r-process elements, the upper bound of $10^{-2}~M_{\odot}$ for the ejected mass from 1.35-1.35~$M_{\odot}$ binaries (i.e for EoSs with small $R_{1.35}$) would be compatible with a merger rate of $4\times 10^{-5}/\mathrm{yr}$. These rate estimates lie in the ballpark of theoretical predictions ranging from $10^{-6}$ to $10^{-3}$ per year~\citep{2010CQGra..27q3001A}. This implies that all EoSs of our survey are compatible with NS mergers being the dominant or a major source of r-process elements.

This conclusion on the merger rate may be tested against future observations, in particular by multiple gravitational-wave detections or frequent observations of electromagnetic counterparts. Our work emphasizes that, in addition to a more accurate merger rate,  information on the high-density EoS is needed to shed light on NS mergers as a major source of r-process elements. More specifically, a merger rate of about $4\times 10^{-5}/\mathrm{yr}$ may imply either that nearly all heavy r-process elements are made by NS mergers in the case of a soft high-density EoS with small NS radius $R_{1.35}$, or that only a tenth of the observed r-process material originates from NS binaries if a stiff NS EoS with large $R_{1.35}$ is confirmed.

Inversely, considering the robustness of the r-process nucleosynthesis in NS mergers, one can infer that, assuming a minimal production of $\sim 10^{-3}~M_{\odot}$ of r-nuclei per event a constant merger rate during the life of the Galaxy, this rate cannot be higher than roughly $4\times 10^{-4}/\mathrm{yr}$; otherwise the galactic  r-process material would be more than presently observed in the Galaxy. This bound is comparable to the ``optimistic'' limit given in~\citet{2010CQGra..27q3001A}. Thus, the r-process element content in the Galaxy establishes further, independent evidence for an upper limit on the merger rate below $\sim 4\times 10^{-4}/\mathrm{yr}$. A restriction to soft EoSs, e.g. from other physical constraints like~\citet{2010PhRvL.105p1102H}, would lead to a smaller upper limit for the event rate of NS mergers.

\begin{table*}
 \begin{ruledtabular}
\caption{\label{tab:nucleo}Nucleosynthesis calculations}
\begin{tabular}{l c c c c c c c c }
\tableline \tableline
Model $M_1$-$M_2$ & $M_{\mathrm{ej}}$ & $t_{\mathrm{peak}}$ & $M_{\mathrm{r-process}}/M_{\mathrm{ej}}$ &  $^{232}$Th/$^{238}$U & $^{232}$Th/$^{235}$U & Th/Eu & $E_{\mathrm{heat}}$ & $f$ \\
& $(10^{-3}~M_{\odot})$ & (d) & & & & & (MeV/A) & \\
\tableline
NL3 1.35-1.35   & 2.09 & 0.171& 0.989 & 1.644  & 1.072  & 0.695 & 3.34 & $1.8\times 10^{-6}$ \\
DD2 1.35-1.35   & 3.07 & 0.189& 0.980 & 1.671  & 1.080  & 0.627 & 3.13 & $1.6\times 10^{-6}$ \\
SFHO 1.35-1.35  & 4.83 & 0.228& 0.991 & 1.642  & 1.039  & 0.579 & 3.17 & $1.4\times 10^{-6}$ \\
NL3 1.2-1.5     & 7.95 & 0.338& 0.964 & 1.670  & 1.112  & 0.714 & 3.36 & $1.3\times 10^{-6}$ \\
DD2 1.2-1.5     & 8.79 & 0.354& 0.986 & 1.697  & 1.109  & 0.658 & 3.11 & $1.2\times 10^{-6}$ \\
SFHO 1.2-1.5    &13.39 & 0.418& 0.974 & 1.685  & 1.085  & 0.543 & 3.12 & $1.1\times 10^{-6}$ \\
BSk21 1.35-1.35 & 3.36 & 0.162& 0.948 & 1.660  & 1.023  & 0.704 & 2.97 & $1.6\times 10^{-6}$ \\
BSk20 1.35-1.35 & 4.68 & 0.195& 0.931 & 1.689  & 1.021  & 0.698 & 2.96 & $1.5\times 10^{-6}$ \\
\tableline
 \end{tabular}
\tablecomments{Selected models for which nucleosynthesis calculations were performed. $M_{\mathrm{ej}}$ is the amount of unbound matter, whereas $t_{\mathrm{peak}}$ is the peak time of an optical transient associated with a NS merger (see Sect.~\ref{sec:opttrans}). The fourth column gives the fraction of the ejecta which is processed into r-process elements. The production ratios of certain elements and isotopes are provided in the fifth to seventh columns. $E_{\mathrm{heat}}$ denotes the total amount of energy released by radioactive decays (without neutrino energy). The factor $f$ approximates the radioactive heat generation around the time of the optical peak luminosity relative to the rest-mass energy of the ejecta (see text).}

 \end{ruledtabular} 
\end{table*}
\subsection{Actinide production ratios and stellar chronometry}
Some of the heaviest long-lived radioactive nuclei produced by the r-process can be used as nucleo-cosmochronometers. In particular the abundance ratios of thorium to europium and thorium to uranium have been proposed for estimating the age of the oldest stars in our Galaxy. More specifically, a simple comparison of the observed abundance ratio with the production ratio can provide an age estimate of the contaminated object~\citep{1987Natur.328..127B,1993A&A...274..821F,1999ApJ...521..194C,1999A&A...346..798G,2001A&A...379.1113G,2001Natur.409..691C,2007ApJ...660L.117F,2008ARA&A..46..241S}. In addition, if we consider low-metallicity stars polluted by a small number of nucleosynthetic events that took place just before the
formation of the stars, the age of the star can be estimated without calling for a complex model of the chemical evolution of the Galaxy. The major difficulty of the methodology is therefore related to the theoretical estimate of the r-production ratio and the corresponding uncertainties of astrophysics and nuclear physics origin that may affect this prediction.

In this respect, the ${}^{232}$Th to ${}^{238}$U chronometry has been shown to be relatively robust, in particular in comparison with the Th/Eu chronometry, i.e to be less affected by the still large astrophysics and nuclear physics uncertainties affecting our understanding of the r-process nucleosynthesis~\citep{1999A&A...346..798G,2001Natur.409..691C,2007ApJ...660L.117F}. In Tab.~\ref{tab:nucleo} we provide the production ratios of the ${}^{232}$Th to ${}^{238}$U isotopes based on our NS merger simulations. Assuming that r-process enhanced metal-poor stars were enriched by one or a few NS merger events, we can derive ages within this scenario. From the observed ratio of $\log{\mathrm{(U/Th)}}_{\mathrm{obs}}=-0.94\pm 0.09$ for the metal-poor star CS31082-001~\citep{2001Natur.409..691C} (see also \citet{2001A&A...379.1113G} for updated values), we compute the age as $ \Delta t=21.8 \left[ \log{\mathrm{(U/Th)}}_{0}-\log{\mathrm{(U/Th)}}_{\mathrm{obs}} \right]=15.7$~Gyr with the production ratio $\log{\mathrm{(U/Th)}}_{0}=-0.22\pm0.01$ (see Tab.~\ref{tab:nucleo}). The observational uncertainty of 0.09 dex in this case dominates the error on the age estimate since it amounts to about 2.0~Gyr, while the theoretical uncertainties associated with the different EoSs (Tab.~\ref{tab:nucleo}) give an 0.2~Gyr error only.

Additional uncertainties stem from the nuclear physics aspects of the r-process nucleosynthesis. In particular, the Th and U production is known to be sensitive to the $\beta$-decay, neutron capture and fission rates adopted in the network calculation. For the specific case of the 1.35-1.35 $M_{\odot}$ merger simulation based on the DD2 EoS, about 20 different abundance calculations based on different nuclear physics ingredients were performed in order to estimate the uncertainties affecting the $^{232}$Th to $^{238}$U production ratio. These inputs include i) different mass models for the estimate of the $\beta$-decay and neutron capture rates, namely the recent microscopic Skyrme \citep{2010PhRvC..82c5804G} and Gogny \citep{2009PhRvL.102x2501G} Hartree-Fock-Bogolyubov (HFB) mass models as well as the macroscopic-microscopic finite-range droplet model (FRDM) \citep{1995ADNDT..59..185M}, ii) different neutron-nucleus optical potentials \citep{1977PhRvC..16...80J,2003NuPhA.713..231K}, iii) different reaction mechanisms, such as the resonant compound or the direct capture models, to estimate the neutron-capture rates \citep{2012PhRvC..86d5801X}, iv) different predictions of the fission barrier heights to estimate the spontaneous, neutron-induced and $\beta$-delayed fission rates, namely the HFB  \citep{2007PhRvC..75f4312G} and Thomas-Fermi \citep{1999PhRvC..60a4606M} fission barriers and v) different prescriptions for the fission fragment distributions \citep{1975NuPhA.239..489K,2010PhRvL.104u2501S}. The final $^{232}$Th to $^{238}$U production ratio is found to lie between 1.41 and 2.30 leading to an age estimate of the metal-poor star CS31082-001 of $14.9 \pm 2.3$~Gyr. The error associated with nuclear physics uncertainties is therefore of the same order as those affecting the observation and significantly larger than those related to the EoS. The largest age of 17.2~Gyr is obtained when use is made of the Gogny D1M mass model \citep{2009PhRvL.102x2501G}, while the smallest estimate of 12.6~Gyr is found with the FRDM mass predictions. The Thomas-Fermi fission barriers \citep{1999PhRvC..60a4606M} also lead to rather larger age with respect to the HFB predictions. More details on the r-process sensitivity to the nuclear physics input within the NS merger model will be given in a forthcoming paper.

The derived age of this halo star lies within the ballpark of other age estimates~\citep{2001Natur.409..691C}. For the metal-poor star HE~1523-0901 ($\log{\mathrm{(U/Th)}}_{obs}=-0.86\pm0.13$)~\citep{2007ApJ...660L.117F} our ${}^{232}$Th to ${}^{238}$U production ratio implies an age of about $14.0\pm2.8$~Gyr, which is also within the range of other calculations~\citep{2007ApJ...660L.117F}.

Tab.~\ref{tab:nucleo} also lists the production ratios of ${}^{232}$Th to ${}^{235}$U as well as Th to Eu. For age estimates the Th/Eu chronometer has been widely used, although it remains highly sensitive to all types of uncertainties. In this case, the stellar age is derived from $ \Delta t=46.7~{\rm Gyr} \left[ \log{\mathrm{(Th/Eu)}}_{0}-\log{\mathrm{(Th/Eu)}}_{\mathrm{obs}} \right]$, so that a 25\% error on the production or observed Th/Eu ratio gives rise to an uncertainty of about 5~Gyr on the stellar age. Special care of the associated uncertainties should therefore be taken when applying this chronometer pair~\citep{2001A&A...379.1113G}. Considering our production ratio of about $\log{\mathrm{(Th/Eu)}}_{\mathrm{obs}}=-0.20\pm 0.06$ (see Tab.~\ref{tab:nucleo}), we find an age of $ \Delta t=17.7\pm2.8$~Gyr for HE~1523-0901 ($\log{\mathrm{(Th/Eu)}}_{\mathrm{obs}}=-0.58$, with an additional 4.8~Gyr uncertainty based on observation~\citep{2007ApJ...660L.117F}). We stress again here that as long as the r-process site remains unidentified, corresponding uncertainties of the production ratios have to be taken into account  as well as uncertainties arising from the nuclear physics input in network calculations. We also point out the possibility that there is a measurable event-to-event variation in the production ratios, for instance in the case of NS mergers caused by the unknown binary configuration (see Tab.~\ref{tab:nucleo}), but also for other sites a progenitor dependence cannot be excluded. 

The production ratios summarized in Tab.~\ref{tab:nucleo} are of particular importance since NS mergers may well be a major source of r-process elements and current supernova models cannot provide suitable conditions for the formation of the heaviest r-process elements~\citep{2008ApJ...676L.127H,2008A&A...485..199J,2010ApJ...722..954R,2010PhRvL.104y1101H,2010A&A...517A..80F,2011ApJ...726L..15W}. The relatively reliable age estimates from the ${}^{232}$Th to ${}^{238}$U ratio are compatible with the age of the universe, and thus NS mergers cannot be excluded as the source of the contamination of the considered metal-poor stars.

\subsection{Nuclear heating}\label{ssec:heat}
Another outcome of our nucleosynthesis calculations is the determination of the heating due to radioactive decays in the ejecta. This is in particular important because the radioactive heating provides the energy source for an optical counterpart associated with NS mergers (see~\citet{1998ApJ...507L..59L},~\citet{2005astro.ph.10256K} and~\citet{2010MNRAS.406.2650M} and Sect.~\ref{sec:opttrans}). Our calculations allow us to check the robustness and general behavior of the heating rate. In all models the heating rate due to radioactive decays (beta-decays, fission and alpha-decays) looks similar (see Fig.~3 in~\citet{2011ApJ...738L..32G}). For instance at the time $t_{\mathrm{peak}}$, when the luminosity of the optical transient reaches its maximum (typically several hours; see Sect.~\ref{sec:opttrans}), the heating rate varies from $3\times 10^{10}$~erg/g/s to $1\times 10^{11}$~erg/g/s for the cases where detailed nucleosynthesis calculations were made. This implies that the heating efficiency $f\equiv \dot{Q}(t_{\mathrm{peak}})t_{\mathrm{peak}}/{M_{\mathrm{ej}}c^2}$ that enters the estimates of optical emission properties (see Sect.~\ref{sec:opttrans} and~\citet{1998ApJ...507L..59L} and~\citet{2010MNRAS.406.2650M}), varies between $1.1\times 10^{-6}$ and  $1.8\times 10^{-6}$ (see Tab.~\ref{tab:nucleo}). One observes a moderate dependence of $f$ on the EoS and the mass ratio. This is caused by the longer duration $t_{\mathrm{peak}}$ of the emission peak for larger ejecta masses in models with soft EoSs or asymmetric binaries. Overall, $f\approx 1.5 \pm 0.3 \times 10^{-6}$ seems to be a fair approximation, which is half of the value suggested in~\citet{2010MNRAS.406.2650M} for a longer peak time of about one day. As detailed in Sect.~\ref{sec:opttrans}, we expect from our relativistic merger simulations shorter peak times in the range of 2 to 7 hours for symmetric binaries. Note that a fraction of the energy of radioactive decays is released in gamma-rays, of which a fraction may escape from the ejecta without efficient thermalization (see~\citet{2005astro.ph.10256K,2010MNRAS.406.2650M}). The assumption of a full thermalization of gamma-rays implies that the factor $f$ is somewhat overestimated. The total amount of energy released by radioactive decays (without neutrinos) is in the range of about $3.2\pm0.2$~MeV per nucleon (see Tab.~\ref{tab:nucleo}).

\section{Optical counterparts and radio remnants}\label{sec:opttrans}
The radioactive decay of the synthesized r-process elements generates heat, which is deposited in the ejecta and powers an optical display~\citep{1998ApJ...507L..59L,2005astro.ph.10256K,2010MNRAS.406.2650M,2011ApJ...736L..21R,2011ApJ...738L..32G}. This electromagnetic counterpart of a NS merger is potentially observable with existing and upcoming optical surveys such as the Palomar Transient Factory, the Synoptic All Sky InfraRed Survey, the Panoramic Survey Telescope and Rapid Response System, and the Large Synoptic Survey Telescope (see e.g.~\citet{2009MNRAS.400.2070S} and~\citet{2012ApJ...746...48M} for a compilation of certain characteristics of these facilities). We also refer to~\citet{2011ApJ...734...96K} for a report on attempts to detect signatures of a radioactively powered transient in the light curve following a gamma-ray burst. The detection of such optical signals would provide valuable information on ejecta properties and the sky position of an event. As detailed in the next section, the peak luminosity, peak time, peak width, and effective temperature depend on the amount of ejecta and the expansion velocity. From an observation one might therefore derive these ejecta characteristics, which are otherwise only accessible by numerical simulations. Detecting radioactively powered emission and determining ejecta masses will consolidate the role of NS mergers for the enrichment of the Galaxy with heavy r-process elements. A precise localization of an unambiguously identified optical transient of a NS merger event will help improving the sensitivity of gravitational-wave detections, and will provide information about the host galaxy or environment. Moreover, if the distance scale of the events can be constrained, better observational limits on the merger rate will become available.
\subsection{Model}
The bolometric peak luminosity of an optical transient associated with a NS coalescence can be estimated by
\begin{equation}\label{eq:Lpeak}
L_{\mathrm{peak}}\approx 5\times 10^{41} \mathrm{erg/s} \left(\frac{f}{10^{-6}} \right)\left(\frac{v}{0.1 c} \right)^{1/2}\left(\frac{M_{\mathrm{ej}}}{10^{-2}M_{\odot}} \right)^{1/2}
\end{equation}
with the average outflow velocity $v$, the ejecta mass $M_{\mathrm{ej}}$ and the heating efficiency $f$ already introduced in Sect.~\ref{ssec:heat}~\citep{1982ApJ...253..785A,1998ApJ...507L..59L,2010MNRAS.406.2650M}.

The time of the peak luminosity and the effective temperature at the time of the maximum luminosity can also be expressed as functions of the outflow velocity, the ejecta mass, and $f$:
\begin{equation}\label{eq:time}
t_{\mathrm{peak}}\approx 0.5~\mathrm{d} \left(\frac{v}{0.1 c} \right)^{-1/2}\left(\frac{M_{\mathrm{ej}}}{10^{-2}M_{\odot}} \right)^{1/2},
\end{equation}
\begin{equation}\label{eq:Tpeak}
T_{\mathrm{peak}}\approx 1.4\times 10^4 \mathrm{K} \left(\frac{f}{10^{-6}} \right)^{1/4} \left(\frac{v}{0.1 c} \right)^{-1/8}\left(\frac{M_{\mathrm{ej}}}{10^{-2}M_{\odot}} \right)^{-1/8}
\end{equation}
(see~\citet{1998ApJ...507L..59L} and~\citet{2010MNRAS.406.2650M}).
Within this model the width $\Delta t_{\mathrm{peak}}$ of the luminosity peak is proportional to $t_{\mathrm{peak}}$~\citep{bookArnett,2009ApJ...703.2205K}. On the basis of the one-zone model in~\citet{2011ApJ...738L..32G} we find that the full width at half maximum can be very well approximated as
\begin{equation}
\Delta t_{\mathrm{peak}} \approx 2.5 t_{\mathrm{peak}}.
\end{equation}
The above formulas can be understood by some general considerations. For larger $M_{\mathrm{ej}}$ or smaller $v$ the ejecta need a longer time to become transparent. If the outflow gets optically thin at a later time, expansion cooling reduces the effective temperature at $L_{\mathrm{peak}}$. Since, however, the increase of the emission radius $R_{\mathrm{peak}}\approx v\times t_{\mathrm{peak}}$ dominates the decrease of the temperature in the Stefan-Boltzmann law, $L_{\mathrm{peak}}\propto T_{\mathrm{peak}}^4 R_{\mathrm{peak}}^2$, the peak luminosity increases with larger $M_{\mathrm{ej}}$ and higher $v$. The simple scaling laws of Eqs.~\ref{eq:Lpeak} to~\ref{eq:Tpeak} for the properties of optical transients are confirmed by calculations in~\citet{2010MNRAS.406.2650M},~\citet{2011ApJ...736L..21R} and~\citet{2011ApJ...738L..32G}.

\subsection{EoS dependence}
\begin{figure*}
\includegraphics[width=8.9cm]{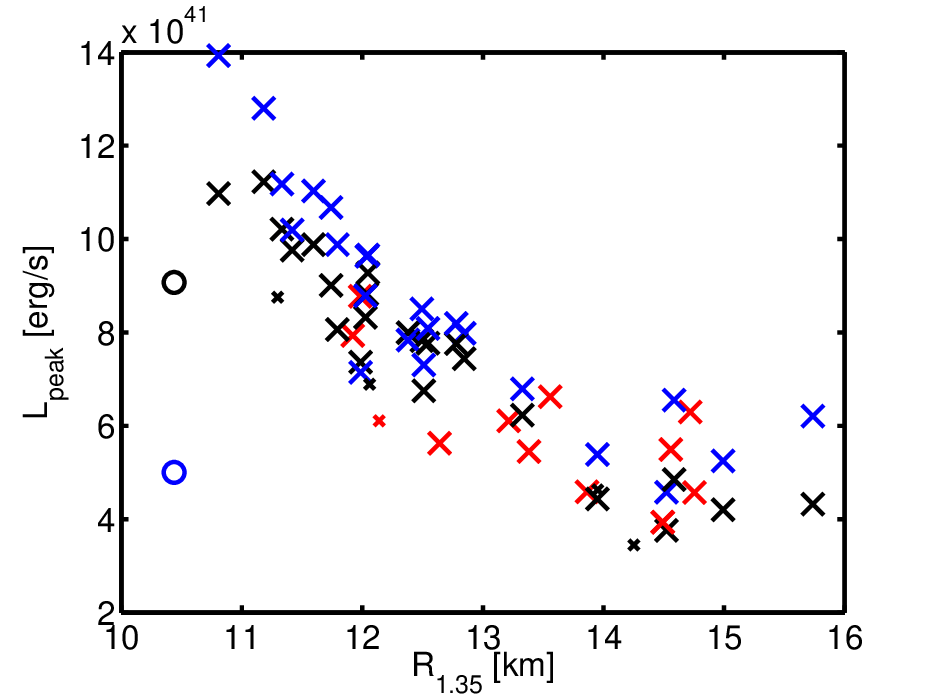}\includegraphics[width=8.9cm]{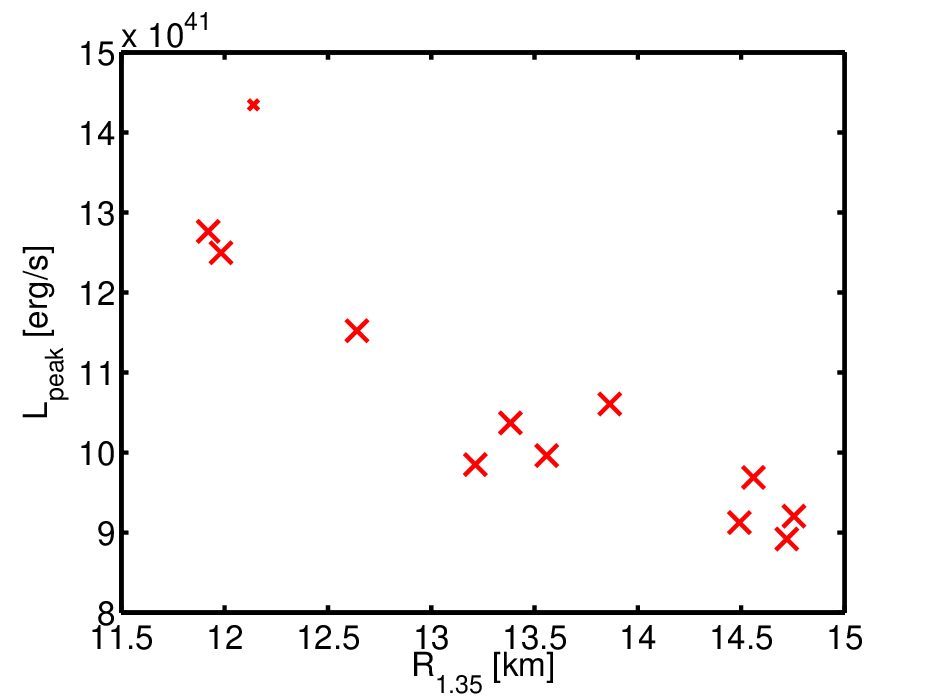}\\
\includegraphics[width=8.9cm]{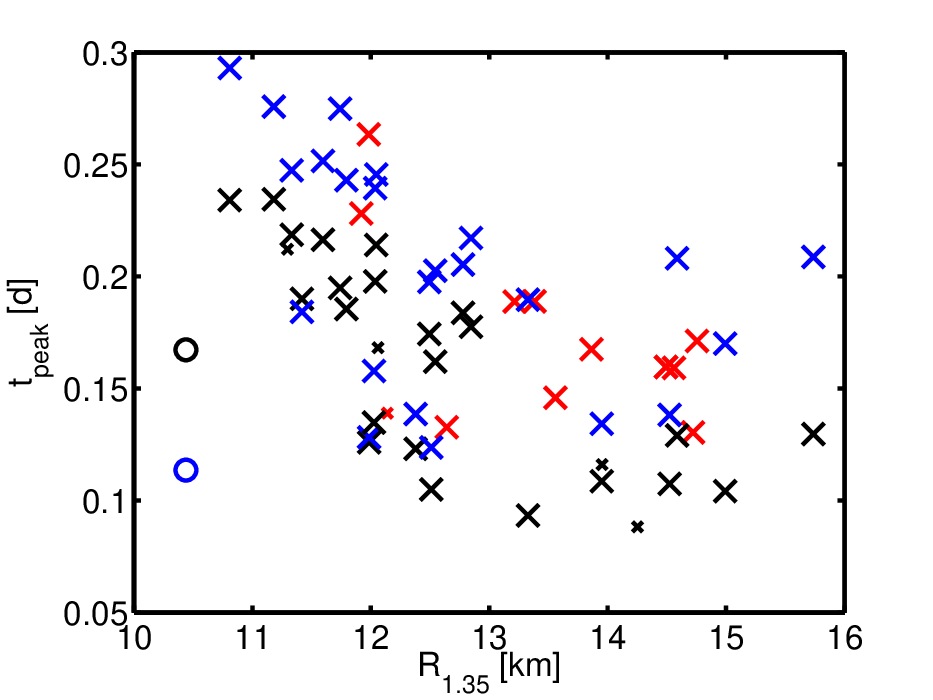}\includegraphics[width=8.9cm]{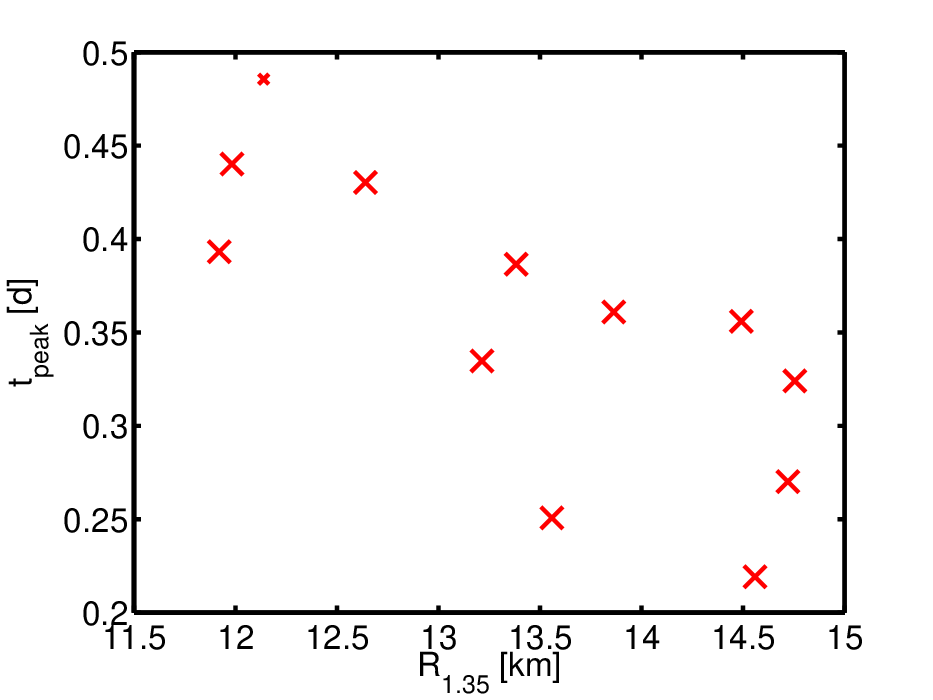}\\
\includegraphics[width=8.9cm]{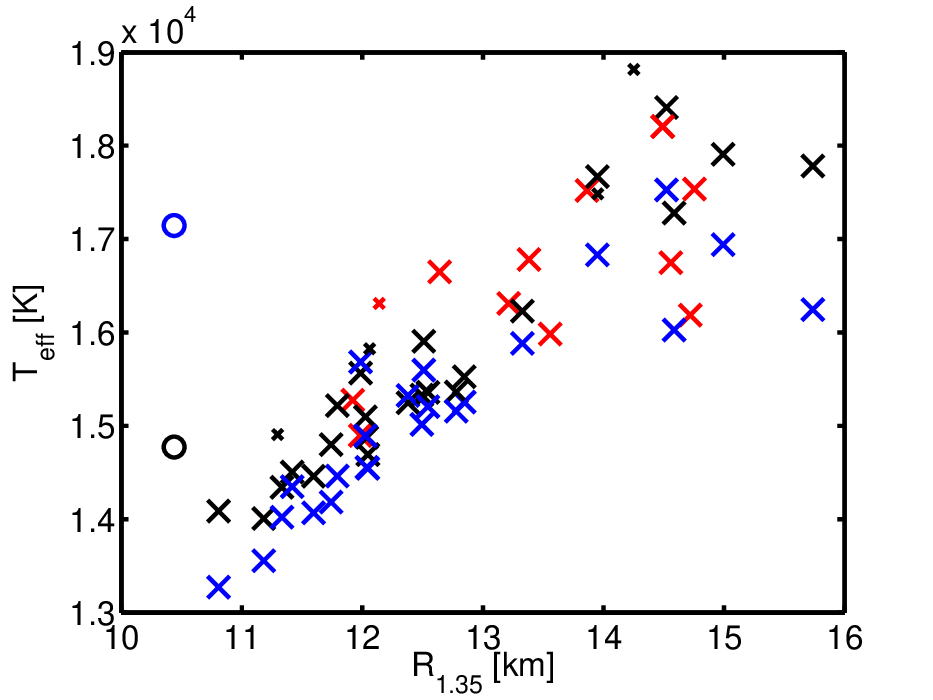}\includegraphics[width=8.9cm]{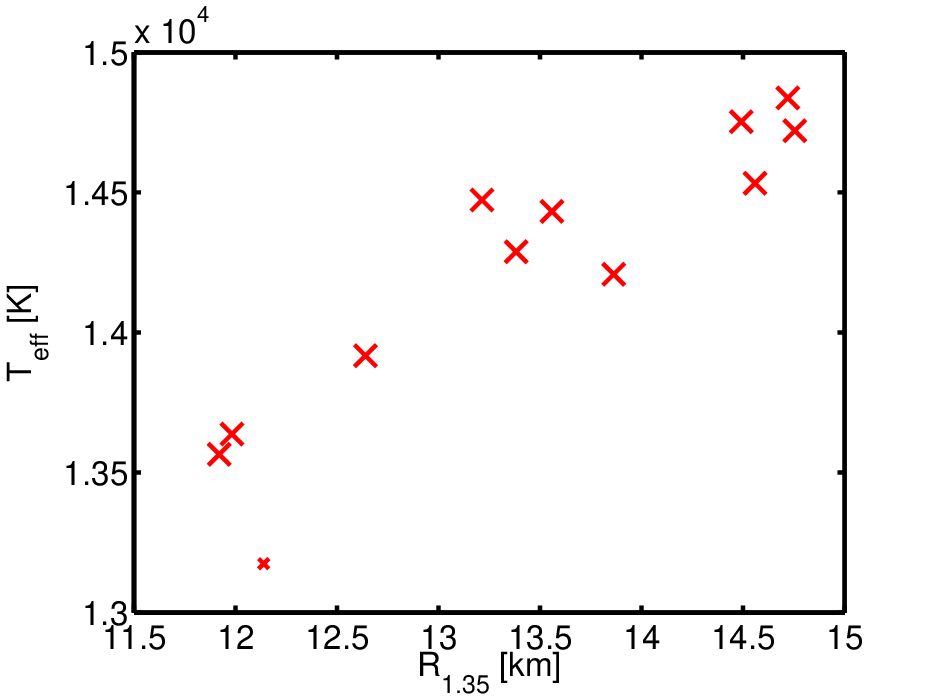}\\
\caption{\label{fig:lpeakr135}Estimated properties of the optical transients for symmetric (1.35-1.35~$M_{\odot}$) mergers (left panels) and asymmetric (1.2-1.5~$M_{\odot}$) mergers (right panels) for different EoSs characterized by the NS radius $R_{1.35}$. The symbols have the same meanings as in Fig.~\ref{fig:mejr135}. The top panels show the bolometric peak luminosity, the middle panels the corresponding peak timescale, and the bottom panels the effective temperature at the time of the peak luminosity.}
\end{figure*}

\begin{figure}
\includegraphics[width=8.9cm]{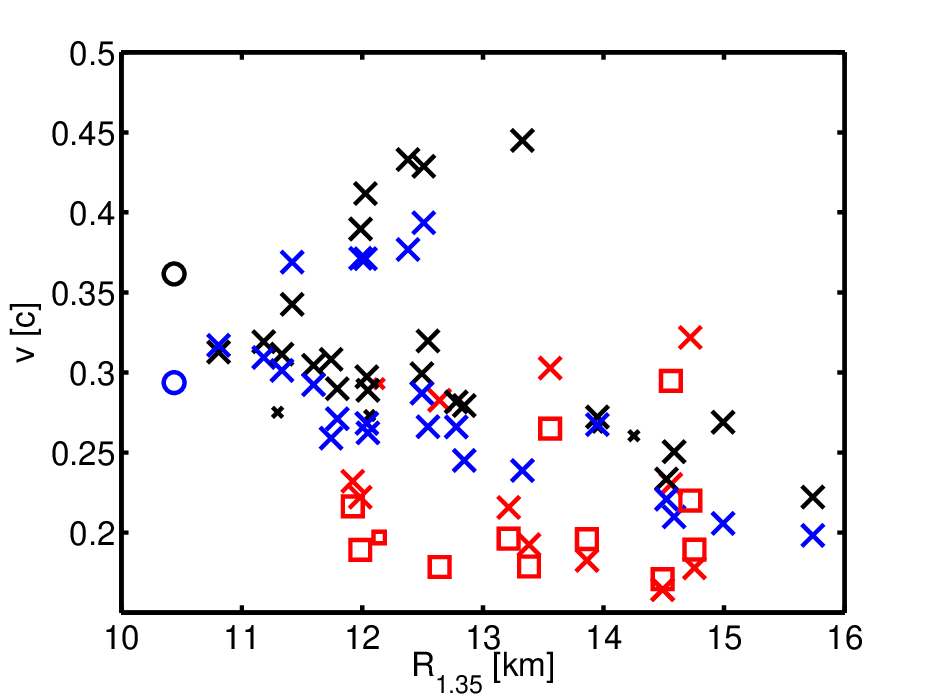}
\caption{\label{fig:v135135}Average ejecta expansion velocity for 1.35-1.35~$M_{\odot}$ mergers (with symbols analog to Fig.~\ref{fig:mejr135}) and for 1.2-1.5~$M_{\odot}$ (red squares) for different EoSs characterized by the corresponding radius $R_{1.35}$ of a nonrotating NS with a mass of 1.35~$M_{\odot}$.}
\end{figure}

The panels on the left side of Fig.~\ref{fig:lpeakr135} display the peak luminosity, the peak time  and the effective temperature of electromagnetic counterparts of 1.35-1.35~$M_{\odot}$ mergers for different EoSs, which are characterized by the radii of the corresponding 1.35~$M_{\odot}$ NSs. A clear dependence of the optical display on the compactness of the NSs can be seen, with soft EoSs yielding brighter transients, which peak on longer timescales with a lower effective temperature. The relatively clear relations are mainly a consequence of the strong EoS impact on $M_{\mathrm{ej}}$ (Fig.~\ref{fig:mejr135}), while the average outflow velocity varies only by a factor of 3 (see Fig.~\ref{fig:v135135} and Table~\ref{tab:models}). The average expansion velocities show the tendency of being higher for EoSs which yield smaller $R_{1.35}$ (see Fig.~\ref{fig:v135135}). This is consistent with the reasoning used before that more compact NSs lead to more violent collisions. However, the relation between the average outflow velocity and $R_{1.35}$ is not very tight. For symmetric binaries the outflow velocities (measured 10~ms after the merging at which time the asymptotic values are fairly well determined) vary (with larger scatter) between 0.16 and 0.45 times the speed of light.

Also for asymmetric 1.2-1.5~$M_{\odot}$ binaries an EoS dependence of optical counterpart properties is observed (right side of Fig.~\ref{fig:lpeakr135}). As in the case of equal-mass mergers, EoSs with smaller $R_{1.35}$ lead to more mass ejection and therefore to more luminous events with longer $t_{\mathrm{peak}}$ and lower $T_{\mathrm{eff}}$. Compared to symmetric binaries asymmetric setups generally produce brighter transients, which reach their peak luminosities on longer timescales and therefore lower effective temperatures, because their ejecta masses are higher whereas the expansion velocities are comparable with those of symmetric systems described by the same EoS (see Fig.~\ref{fig:v135135}). For all considered quantities the asymmetric models exhibit a milder EoS dependence. Plotting the peak luminosity as a function of the binary mass ratio and the total binary mass for the NL3, DD2 and SFHO EoSs reveals a qualitatively similar behavior as the ejecta masses shown in Fig.~\ref{fig:binpara3eos}.

The relations shown in Fig.~\ref{fig:lpeakr135} suggest the possibility to constrain NS radii and thus the high-density EoSs from observations of optical transients associated with NS mergers. Optimally such a detection could be supplemented by a gravitational-wave measurement, which provides the involved binary masses, the distance, and the merger time. But even without an associated gravitational-wave signal the observable features of a transient may have the potential to yield constraints for the NS EoS. The combinations of $L_{\mathrm{peak}},~t_{\mathrm{peak}}$, and $T_{\mathrm{eff}}$ vary systematically with the NS properties. For instance a low peak luminosity, a small peak width, and a high effective temperature imply a large NS radius.
\subsection{Implications for observations}
Symmetric 1.35-1.35~$M_{\odot}$~binaries are predicted to be the most common configurations and thus are likely to be the ones first and most frequently observed. Unfortunately, the 1.35-1.35~$M_{\odot}$~systems yield the smallest ejecta masses and thus the lowest luminosities and peak timescales. The peak widths are important to estimate the prospects of blind searches, whereas the peak time sets the scale for the response time after a gravitational-wave trigger.

The nearly complete coverage of EoS possibilities by our survey allows us to determine the possible range of signal properties of optical transients associated with NS merger events. From our survey we find that the optical peak luminosity of a 1.35-1.35~$M_{\odot}$~NS merger should be expected to be between about $3\times 10^{41}$~erg/s and $14\times 10^{41}$~erg/s corresponding to absolute bolometric magnitudes of $M=-15.0$ and $M=-16.7$. The peak times range from only 2~hours to 7~hours, and the duration of the emission is expected to be between 4.8~hours and 18~hours depending on the high-density EoS. Note that the ballpark of our models yields fainter and shorter transients than typical estimates based on Newtonian models, which in general obtain higher ejecta masses and lower average expansion velocities~\citep{2011ApJ...736L..21R,2012arXiv1206.2379K,2012arXiv1204.6240R,2012arXiv1204.6242P,2012arXiv1210.6549R} (see also the discussion in Sect.~\ref{sec:masses}). While for the peak luminosity these differences lead to partially compensating effects, the timescales are more strongly affected. For symmetric binaries even the maximum peak time of about 7~hours found in our sample is well below most predictions based on Newtonian models~\citep{2010MNRAS.406.2650M,2011ApJ...736L..21R,2012arXiv1204.6242P,2012arXiv1204.6240R}.

As shown in Fig.~\ref{fig:ejectageo}, during the early stages of the expansion the ejecta exhibit a fair asymmetry between polar and equatorial directions. The forumlas used in this section for estimating the properties of optical counterparts assume spherically symmetric outflows. It remains to be explored whether the donut-like shape visible in Fig.~\ref{fig:ejectageo} persists at late times or whether the outflow becomes more symmetric when the peak of the optical display occurs. Multidimensional radiation transport calculations coupled to long-term hydrodynamical simulations are required to determine to which extent the simplified emission model provides reliable estimates of the observable properties of the electromagnetic transients in dependence on the observer direction.

The latest detailed atomic models~\citep{2013arXiv1303.5788K,2013arXiv1303.5787B} predict that the opacity of r-process elements may be significantly enhanced compared to iron group elements, whose opacities have been adopted in Eqs.~\eqref{eq:Lpeak} to~\eqref{eq:Tpeak}. The effects of an increased opacity can be readily estimated because the opacity $\kappa$ enters the prefactors as $\kappa^{-1/2}$ in Eq.~\eqref{eq:Lpeak}, as $\kappa^{1/2}$ in Eq.~\eqref{eq:time}, and as $\kappa^{-3/8}$ in Eq.~\eqref{eq:Tpeak} (see~\citet{2010MNRAS.406.2650M}). While the peak luminosity of the transient is reduced and its duration stretched, the differences between different EoSs remain. By observing two signal features like for instance the peak luminosity and the peak width, one can in principle remove the degeneracy due to the uncertain opacity because every observable individually exhibits a specific approximate correlation with $R_{1.35}$, where the relation is known except for a constant factor.

\subsection{Radio flares}
\begin{figure}
\includegraphics[width=8.9cm]{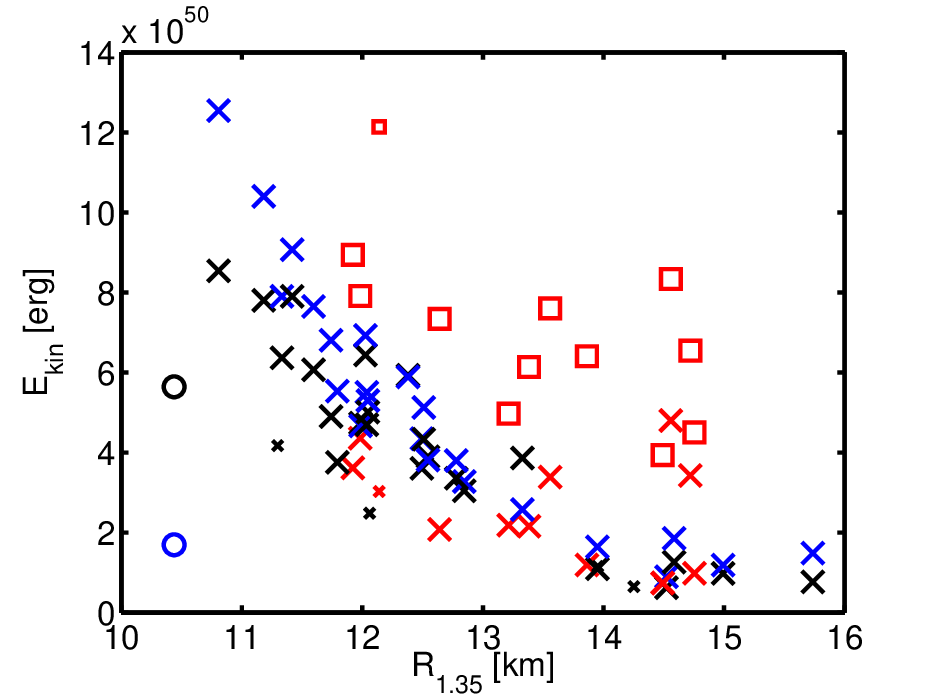}
\caption{\label{fig:ekin}Kinetic energy of the ejecta for 1.35-1.35~$M_{\odot}$ mergers (with symbols analog to Fig.~\ref{fig:mejr135}) and for 1.2-1.5~$M_{\odot}$ (red squares) for different EoSs characterized by the corresponding radius $R_{1.35}$ of a nonrotating NS with a mass of 1.35~$M_{\odot}$.}
\end{figure}

Another potentially observable phenomenon connected with NS mergers is radio emission that is produced by the interaction of the outflowing material with the ambient medium~\citep{2011Natur.478...82N,2012arXiv1204.6242P,2012arXiv1204.6240R}. The ejecta properties which determine the appearance of a radio remnant are the kinetic energy and the outflow velocity. The peak flux density was computed to be proportional to the total kinetic energy $E_{\mathrm{kin}}$ of the outflow and to the (initial) outflow velocity $v$ to a power of about 2.5~\citep{2012arXiv1204.6242P}. For symmetric mergers we find values for the kinetic energy between $6\times 10^{49}$~erg and $10^{51}$~erg (Fig.~\ref{fig:ekin}) and Table~\ref{tab:models}, whereas the average outflow velocities vary from 0.16 to 0.45 times the speed of light (Fig.~\ref{fig:v135135}). The kinetic energy scales well with the expansion velocity, i.e. the models with the highest outflow velocities yield also the highest kinetic energies, and configurations with smaller $v$ result in lower kinetic energies. This implies that the peak flux of radio remnants is uncertain by a factor of 200 because of the variations in the kinetic energy and expansion velocity associated with the incomplete knowledge of the NS EoS. Similar values are found for asymmetric mergers. Moreover, the circumburst densities affecting the signal brightness and length are uncertain, and low densities are likely to reduce the radio detectability~\citep{2005Natur.438..988B,2006ApJ...650..261S,2012ApJ...746...48M,2012ApJ...756..189F,2013arXiv1302.3221F}.

\section{Summary and conclusions}\label{sec:sum}
We have performed relativistic hydrodynamical simulations of NS mergers to investigate the mass ejection, the nucleosynthesis outcome, and the properties of associated optical transients. The main goal of this study was to explore the systematics of the EoS dependence of these aspects by employing a large set of candidate EoSs, while focussing mostly on binaries with two 1.35~$M_{\odot}$ NSs and on asymmetric 1.2-1.5~$M_{\odot}$~systems. 

The unbound ejecta mass is strongly affected by the adopted EoS. We find that the NS compactness is the crucial EoS parameter determining the ejecta properties. Using the radius $R_{1.35}$ of a nonrotating 1.35~$M_{\odot}$ NS to characterize different EoSs we find that ``softer'' EoSs which yield more compact NSs tend to produce more ejecta. The ejecta masses are between $10^{-3}~M_{\odot}$ and $1.5\times 10^{-2}~M_{\odot}$ depending on the EoS and binary mass ratio (Figs.~\ref{fig:mejr135} and~\ref{fig:binpara3eos}). Most of the unbound material originates from the contact interface between the colliding stars for symmetric as well as asymmetric binaries. In the latter case about 25 per cent of the mass ejection are shed off the outer end of the spiral arm like tail into which the lower-mass component is stretched during its final approach to collision with the more massive companion.

A qualitative and quantitative agreement of our SPH simulations (employing the conformal flatness approximation) with fully relativistic grid-based simulations is found, whereas considerable differences compared to Newtonian models concerning the origin and the amount of ejecta are observed. The pronounced spiral arms, for example, which form in Newtonian simulations during the merging of symmetric binaries and whose mass stripping dominates the ejecta, are absent in relativistic mergers of equal-mass NSs. Newtonian models therefore tend to produce considerably higher ejecta masses.

When temperature effects are mimicked by adding a thermal ideal-gas component with a constant ideal-gas index $\Gamma_{\mathrm{th}}$ to EoS models which are provided as zero-temperature barotropes, the best match of the ejecta masses with fully consistent calculations is achieved for a relatively low value of $\Gamma_{\mathrm{th}}=1.5$. This is in conflict with values of 1.8 or 2 which have been widely used and work well for gravitational-wave determinations~\citep{2012PhRvD..86f3001B,2010PhRvD..82h4043B}. This can be understood from the fact that the gravitational-wave signal is produced by the bulk of the merger mass in the high-density regime, whereas the ejecta depend on the thermodynamics of lower-density matter expanding away from the colliding stars and being accelerated by pressure forces.

The binary parameters have qualitatively the same influence on the ejecta masses for all investigated EoSs. A larger binary mass asymmetry leads to a strong increase of the mass of unbound matter, whereas a higher total binary mass results in larger ejecta masses for asymmetric mergers but only a weak increase of $M_{\mathrm{ej}}$ for symmetric systems (Fig.~\ref{fig:binpara3eos}). The occurrence of a prompt collapse of the merger remnant is associated with a significant drop in the ejecta mass. For a given EoS (and no prompt collapse) the smallest amount of ejecta is to a good approximation produced by 1.35-1.35~$M_{\odot}$ binaries, while the ejecta mass can be up to a factor ten higher for other binary configurations. For soft EoSs the ejecta masses show steeper gradients in the binary parameter space ($q,~M_{\mathrm{tot}}$).

Nuclear network calculations show that r-process elements with mass numbers above 130 are robustly produced in the ejecta of NS mergers for all investigated EoSs. The vast majority of ejected material is fission recycled producing a final mass-integrated abundance pattern that resembles closely the solar composition of r-process elements. The robustness with respect to variations of the high-density EoS confirms that NS mergers are a very promising source of r-process elements.

For some EoS models which do not provide the temperature dependence consistently, the hydrodynamical simulations are based on an approximate treatment of thermal effects and the nucleosynthesis calculations rely on an estimation of the temperature in the ejecta. Also in these cases r-process elements are produced with solar distribution, showing the insensitivity to the exact temperature value at the beginning of the network calculations. Moreover, the abundance patterns are not affected by asymmetries in the binary setup. The amazing insensitivity of the outcome of the nucleosynthesis processes can be understood as a consequence of the large neutron excess in the ejected inner-crust material of the merging NSs, which allows for fission recycling in essentially all considered conditions.

By folding with the binary population we identify the results of 1.35-1.35~$M_{\odot}$ mergers as a good approximation for the average ejecta mass per merger event. The main uncertainty in the average ejecta mass per merger is therefore associated with the incomplete knowledge of the NS EoS, which implies variations in the average ejecta mass of about a factor of 10. The observed abundance of r-process matter in the inventory of our Galaxy can be accounted for with a NS merger rate that is compatible with current predictions based on population synthesis and pulsar observations~\citep{2010CQGra..27q3001A}. Final conclusions, however, require a more precise determination of the merger rate, e.g. by gravitational-wave detections or observations of electromagnetic counterparts. Our work implies that in addition to more accurate information on the merger rate also better constraints on the high-density EoS are needed to decide whether NS mergers are a major (or the dominant) source of r-process elements.

Moreover, our simulations in combination with estimates of the Galactic r-process material provide independent evidence that the Galactic merger rate cannot be higher than approximately $4\times 10^{-4}$ events per year if NS mergers should not overproduce heavy r-nuclei compared to observations.

The nucleosynthesis calculations of our survey also provide important information on the production ratios of certain isotopes which are used for nucleocosmochronometry. For instance the ratio of 232-thorium to 238-uranium is found to be about 1.65 with only small variations depending on the high-density EoS and the binary configuration (Tab.~\ref{tab:nucleo}). Using this result we derived ages of metal-poor stars which are consistent with other age estimates. This implies that NS mergers are not excluded as r-process element sources for metal-poor stars and the production ratios provided here for the first time in the NS merger context should be taken into account in stellar age estimates as long as mergers cannot be excluded as r-process sites in the early Galactic history.

Just as the ejecta masses of binary NS mergers exhibit a strong sensitivity to the properties of the nuclear EoS and thus to the radius $R_{1.35}$ of the merging stars, we also predict the optical transients powered by the radioactive energy release in the ejecta to depend on the compactness of the binary components. EoSs which lead to smaller NS radii produce more ejecta and therefore cause brighter optical counterparts, which peak on longer timescales with longer durations and with lower effective temperatures. On the basis of our extensive survey of EoSs, which suggests clear correlations between observable features (luminosity, peak timescale, effective temperature) and NS radii we propose that optical observations of transients associated with NS mergers could yield valuable constraints on the NS EoS.

The very broad range of possibilities included in our EoS sample allows us to bracket the expected range of signal features of optical counterparts associated with NS mergers. Optical transients of 1.35-1.35~$M_{\odot}$ mergers should (at least) reach an absolute bolometric peak magnitude between -15.0 and -16.7 ($3 \times 10^{41}$~erg/s and $14 \times 10^{41}$~erg/s). Depending on the high-density EoS the peak times vary from 2 to 7 hours, implying durations of about 4 to 18~hours, whereas effective temperatures between $1.3\times 10^4$~K and $1.9\times 10^4$~K can be expected. We emphasize that the peak luminosities, peak times, and peak widths of the optical counterparts are found to be considerably lower in our analysis compared to earlier investigations based on Newtonian models~\citep{2010MNRAS.406.2650M,2011ApJ...736L..21R,2012arXiv1204.6242P,2012arXiv1204.6240R}. The reduction is a consequence of the smaller ejecta masses especially for symmetric binaries. Because of the shorter peak time and duration of the optical transient suggested by relativistic merger results, we also find a smaller fraction $f$ of radioactive decay energy relative to the rest-mass energy of the ejecta. While Newtonian merger models yield $f\sim 3\times 10^{-6}$ at the time of the luminosity peak~\citep{2010MNRAS.406.2650M} we obtain $f\sim 1.5\times 10^{-6}$ with little sensitivity to the EoS and the binary parameters.

For different EoSs considerable differences are also found in the properties of radio remnants. Based on our sample of models we estimate an uncertainty of up to a factor of 200 for the theoretical predictions of the brightness of these events.

Future work should address a variety of issues. The hydrodynamical models of NS mergers should include magnetic fields and the effects of neutrino interactions should be explored. Our results also need to be confirmed by fully relativistic merger simulations. The properties of emitted electromagnetic radiation should be computed for detailed multi-dimensional outflow models including the corresponding nuclear network calculations to determine the composition and heating. Radiative transfer calculations will have to be performed to study the observational appearance of the potentially anisotropic ejecta dependent on the viewing direction. This will require employing appropriate opacities of the r-processed material as recently derived by~\citet{2013arXiv1303.5788K}. The corresponding opacity increase by orders of magnitude leads to considerably lower peak luminosities and longer peak timescales than estimated in our work. However, in view of the faster expansion timescales and the lower ejecta masses of our relativistic models the very long durations of radioactive transients calculated by~\citet{2013arXiv1303.5787B} (on the basis of assumed outflow properties) appear to be on the extreme side. In this context it will also be important to determine the contribution of mass ejection from the secular evolution of the merger remnant (a black hole-torus system or hypermassive NS), which will lose mass through neutrino-driven and magnetohydrodynamical outflows. The corresponding matter will increase the ejecta mass (see e.g.~\citet{2013arXiv1304.6720F} for longterm evolution models) and will ultimately have to be taken into account for reliable predictions of the properties of electromagnetic counterparts of NS mergers. However, the differences in the composition and velocities of this additional mass outflow are predicted to lead to a second light curve peak that can be observationally discriminated from the signature of the dynamical ejecta~\citep{2013arXiv1303.5787B}. Therefore the potential of inferring EoS information from the counterpart of the dynamical ejecta is likely to remain unaffected. The robustness of the nucleosynthesis outcome has to be explored concerning variations connected to uncertainties of the nuclear reaction rates. Further work is also needed to address how NS mergers as r-process sources fit into chemical evolution scenarios of the Milky Way, which should explain the observations of r-element enhanced metal-poor stars. Finally, the capabilities of various observational facilities have to be evaluated in view of the bounds on the observable features set by our survey.

\begin{acknowledgments}
We thank Matthias Hempel for helpful discussions and for providing his EoS tables. This work was supported by Sonderforschungsbereich Transregio 7 ``Gravitational Wave Astronomy", and the Cluster of Excellence EXC 153 ``Origin and Structure of the Universe" of the Deutsche Forschungsgemeinschaft, and by CompStar, a research networking programme of the European Science Foundation. S.G is F.R.S.-FNRS Research Associate. Computing resources provided by the Rechenzentrum Garching of the Max-Planck-Gesellschaft and Leibniz-Rechenzentrum Garching are acknowledged.
\end{acknowledgments}


\end{document}